\documentclass[aps,nofootinbib,prb,reprint,superscriptaddress,longbibliography]{revtex4-2}
\usepackage{amsmath, amssymb, bm, braket, circuitikz, derivative,  lipsum, lmodern, pgfplots, tikz, verbatim} %
\pgfplotsset{compat=1.18}
\usepackage[colorlinks,linkcolor=blue,urlcolor=blue,citecolor=blue]{hyperref}
\newcommand{\rd}{\mathrm{d}}

\newcommand{\ZG}{\color{red}}
\newcommand{\bea}{\begin{eqnarray}}
\newcommand{\eea}{\end{eqnarray}}

\begin{document}

\title{Capturing Open Quantum System Dynamics with Compact Dressed Hamiltonians}

\author{Zachary Giles}
\email{z.giles@mail.utoronto.ca}
\affiliation{Department of Physics and Centre for Quantum Information and Quantum Control, University of Toronto, 60 Saint George St., Toronto, Ontario, M5S 1A7, Canada}

\author{Dvira Segal}
\email{dvira.segal@utoronto.ca}
\affiliation{Department of Chemistry, University of Toronto, 80 Saint George St., Toronto, Ontario, M5S 3H6, Canada}
\affiliation{Department of Physics and Centre for Quantum Information and Quantum Control, University of Toronto, 60 Saint George St., Toronto, Ontario, M5S 1A7, Canada}

\date{\today}
\begin{abstract}
We advance the compact dressed Hamiltonian (CDH) framework to simulate the dynamics of open quantum systems strongly coupled to structured bosonic environments. The CDH approach combines reaction-coordinate mapping with a polaron transformation, incorporating strong system-environment hybridization into a dressed representation before truncating the collective bosonic mode. As a result of this truncation, dynamical features are captured with substantially fewer states than in the original representation. The residual environment is treated using a weak-coupling Markovian quantum master equation.
We benchmark the method for spin-boson-type models including a single spin, coupled spins, and spin ensembles, considering both resonant and off-resonant regimes and weak to strong system-bath coupling. 
Comparisons with hierarchical equations of motion and converged reaction-coordinate calculations demonstrate that the CDH representation systematically approaches the reference dynamics with a compact Hilbert space, accurately reproducing oscillation frequencies, relaxation behavior, and long-time populations over a broad parameter regime. The method further captures cooperative phenomena, including bath-mediated interactions, synchronization between spins, and signatures for collective emission. 
Beyond efficient time propagation, the CDH framework provides direct access to the spectrum and eigenmodes of the Liouvillian. We demonstrate that slow dynamical modes converge rapidly in the CDH representation, enabling identification of relaxation timescales without fitting time-dependent observables. As a whole, the CDH framework provides an efficient and physical description of strongly coupled open quantum-system dynamics. 
\end{abstract}

\maketitle

\section{Introduction}

The strong coupling between a quantum system and its environment is central to many condensed-phase dynamical processes \cite{nitzanChemical2006,ishizakiAdequacy2009,ishizakiQuantum2012,xuNoncanonical2016}. In electron transfer, exciton dynamics, and energy transport in molecular and biomolecular systems, electronic degrees of freedom may interact strongly with intra- or inter-molecular nuclear vibrations, solvent fluctuations, and other environmental modes. These interactions can substantially renormalize system energetics, affect relaxation, decoherence, and transport pathways, and thus alter reaction dynamics.
Describing such regimes requires approaches that go beyond the conventional weak-coupling treatment of the environment and explicitly account for the interplay between coherent quantum dynamics and strong environmental interactions.

Beyond electron-vibration coupling, ultrastrong cavity coupling to atomic and molecular degrees of freedom has been realized across a wide range of physical platforms. Strong coupling of cavity modes to both electronic and vibrational excitations in molecules was observed \cite{ebbesenIntroduction2023, liMolecularPolaritonicsChemical2022a, colesStrong2014, dongManipulation2024, sandikCavityenhanced2025}, while analogous regimes have been reached in quantum materials \cite{schlawinCavity2022, muellerDeep2020} and in superconducting qubits coupled to resonators \cite{majerCoupling2007, friskkockumUltrastrong2019, forn-diazUltrastrong2019}. These light-matter interactions have been exploited to control transport processes \cite{dongManipulation2024}, modify chemical reactivity \cite{ebbesenIntroduction2023, liMolecularPolaritonicsChemical2022a}, engineer collective phenomena \cite{schlawinCavity2022}, and offer promising avenues for fast and efficient quantum information processing \cite{friskkockumUltrastrong2019}.

With growing access to the strong coupling regime, there is a need for both accurate and efficient theoretical and numerical methods to study open quantum systems in this regime. This remains challenging as conventional-convenient perturbative approaches break down under strong system-environment interactions \cite{leboiteTheoretical2020,nitzanChemical2006}. 
Polaron-frame quantum master equations are applicable in the strong coupling limit, but suffer from other limitations such as being restricted to the nonadiabatic (small tunneling) regime \cite{Silbey1984Variational,
Segal2002Conduction,
Kolli2011Polaron,
Ahsan24}. 
Numerically exact techniques such as hierarchical equations of motion (HEOM) \cite{tanimuraNumerically2020, katoHierarchical2018, suessHierarchy2014, chenBexcitonics2024,FrancoTenso25,Thoss24A,Thoss26} and quantum path-integral approaches \cite{makriTensor1995, Makri20, strathearnEfficientNonMarkovianQuantum2018b,SegalPI10,Simine13,chenInchworm2017,David23,Guy15,goulkoTransient2025}
provide accurate benchmarks, but are computationally demanding and typically offer limited analytical insight into the underlying dissipative dynamics.
An ideal approach would retain the efficiency and interpretability of analytical methods while accurately capturing key dynamical properties, such as relaxation and decoherence rates, steady-state populations, and dominant dissipative pathways.

\begin{figure*}[ht]
    \centering
\includegraphics[width=0.8\linewidth]{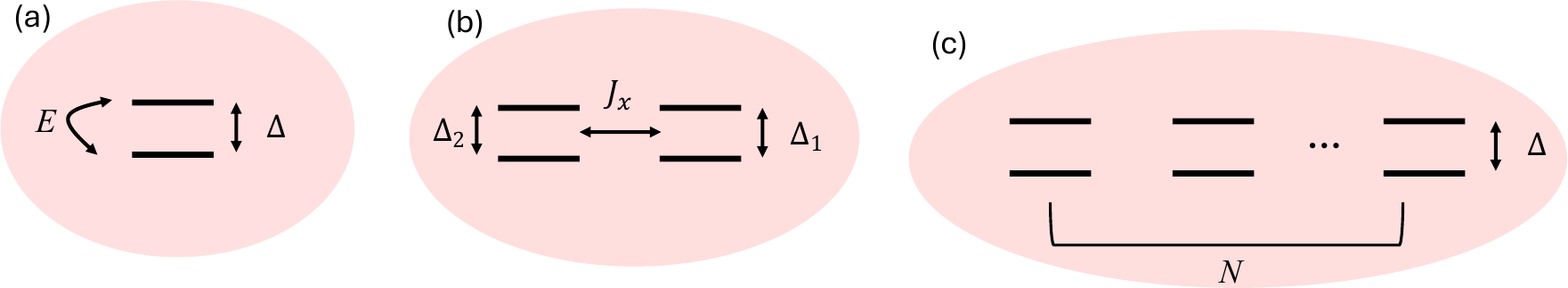}
    \caption{Diagram of models studied in this paper. (a) The spin-boson model with a spin-1/2 system, with level splitting $\Delta$ and tunneling $E$, coupled to a thermal bath (colored background). 
    (b) Two spins with energy splittings $\Delta_1$ and $\Delta_2$ coupled to common bath and to each other with strength $J_x$. 
    (c) $N$ identical spins with energy difference $\Delta$ coupled to a common bath.}
    \label{fig:systems}
\end{figure*}

The recently developed compact cavity-dressed Hamiltonian (CDH) method \cite{garwolaCompact2026} provides a promising route toward this goal. The CDH method belongs to the broader class of Markovian embedding, in particular pseudomode or auxiliary mode methods which extract modes and add them to the system \cite{garrawayNonperturbative1997, lambertModelling2019, pleasanceGeneralized2020, Marlon20, ben-asherMemory2025, huangCoupled2026}. In a similar spirit, the effective Hamiltonian approach \cite{anto-sztrikacsEffective2023, anto-sztrikacsEffectiveHamiltonian2023},  
combines a reaction-coordinate (RC) mapping
\cite{gargEffect1985, iles-smithEnvironmental2014a, nazirReaction2018, correaPushing2019}
with a polaron transformation to incorporate 
a collective environmental mode that captures the dominant system-environment coupling
into an effective system Hamiltonian. The transformed Hamiltonian is subsequently truncated, yielding a compact low-dimensional description while preserving the essential effects of strong system-bath coupling. The RC mode, which couples to both the system and the residual bath, can also represent a leaky cavity mode, with its frequency set either in resonance with Bohr's frequencies of the system, or off-resonant.

Earlier implementations of the Effective Hamiltonian method were restricted to truncations retaining only the RC ground state, which required this frequency to be a large energy scale and thus excluded resonant configurations \cite{anto-sztrikacsEffective2023, anto-sztrikacsEffectiveHamiltonian2023}.
The CDH method \cite{garwolaCompact2026} overcomes this limitation by systematically including excited RC states in the truncation, substantially extending its range of validity into resonant and near-resonant regimes.
Moreover, because the effective Hamiltonian is obtained analytically, the CDH method provides physical insight into strong-coupling phenomena, including energy-level renormalization and bath-mediated interactions.
While Ref. \cite{garwolaCompact2026}, which introduced the method, focused on ground state properties of the closed system, the present work applies the CDH mapping method to study open quantum dynamics.


In this work, the CDH framework is applied to investigate dissipative dynamics in spin-boson-type models, diagrammed in Fig. \ref{fig:systems}, including single spin, coupled spin systems, and spin ensemble interacting with a thermal environment. Computationally, by benchmarking against numerically exact hierarchical equations of motion simulations, we demonstrate that the CDH representation accurately reproduces the dynamics while offering a significant computational advantage over both the conventional reaction-coordinate approach and numerically exact simulations.  

Beyond efficient dynamical simulations, the CDH framework provides physical information through access to the Liouvillian governing the dissipative dynamics, formulated here within the Redfield approximation. Analysis of the Liouvillian spectrum yields the relaxation rates and the corresponding eigenmodes that govern the transient dynamics and approach to the steady state, providing direct access into the underlying dissipative processes, including a direct means to define relaxation rates without fitting procedures. 

The paper is organized as follows. We describe the CDH mapping method in Sec. \ref{sec:mapp}. Applications to spin-boson models: single spin, two-spin synchronization, and signatures of collective effects in spin chains are presented in Sec. \ref{sec:multspin}. We summarize and discuss future work in Sec. \ref{sec:summ}.


\section{Hamiltonian Dressing Method}
\label{sec:mapp}

We present here the CDH dressing method: it takes an open quantum system Hamiltonian of dimensionality $N\times N$, with a bilinear coupling to a bosonic bath, and prepares a reaction coordinate (also termed collective-coordinate) dressed Hamiltonian, which we refer to as the cavity dressed Hamiltonian of dimension $(M\times N)\times(M\times N)$, weakly coupled to a residual bosonic heat bath.  
Here, $M=1,2,...$ is the degree of truncation of the collective RC mode. 
The CDH method is so-far limited to describe harmonic baths and systems bilinearly coupled to the bath. The method is most suitable for describing baths with a spectral density function that is peaked around a central frequency; here, we use the Brownian function as an example.

The CDH mapping method is an example of a Markovian embedding technique, based on the effective Hamiltonian method developed in Refs. \cite{anto-sztrikacsEffective2023,anto-sztrikacsEffectiveHamiltonian2023}. Starting with an open system Hamiltonian, the mapping includes three steps: (i) A collective coordinate is extracted from the harmonic bath.
This mode is harmonic (and in the light-matter scenario, we refer to it as a ``cavity mode"). It is constructed to make the external system-environment coupling an internal parameter of the extended system+cavity mode. The collective mode in turn couples--potentially weakly--to a residual bath.
(ii) The extended system goes through a polaron transformation to create a hybrid matter-RC Hamiltonian. It serves as a more accurate representation for the combined system.
(iii) The Hilbert space of the extended system (which is infinite for the system+collective mode) is truncated.
In Ref. \cite{anto-sztrikacsEffectiveHamiltonian2023}, the collective mode was truncated to its ground state, essentially recovering an effective Hamiltonian with the same dimension $N\times N$ as the original system. Such an extreme truncation can be justified in the off-resonant regime when the bath's central mode (which essentially becomes the frequency of the RC) is of high frequency relative to transitions in the system.

This procedure was used in several studies to examine the equilibrium \cite{anto-sztrikacsEffective2023}, steady state \cite{anto-sztrikacsEffectiveHamiltonian2023,Brett24}, and dissipative dynamics \cite{garwolaOpen2024} of open quantum systems. 
For the spin-boson model in particular, although it was proved in Ref. \cite{anto-sztrikacsEffective2023} that under this strong truncation of the cavity the method still exactly describes the weak and deep strong coupling limit of the model, deviations emerged at the intermediate coupling limit for both equilibrium and dynamical studies.
In particular, the dynamics reached by enacting complete truncation of the mode ($M=1$) recovered the correct long-time rate constant, but missed dynamical signatures at short times.

The basic mapping method was extended in Ref. \cite{garwolaCompact2026}, to maintain $M$ levels in the collective mode, thus creating what we term a ``compact cavity dressed Hamiltonian". To facilitate analytical evaluations, the polaron transformed Hamiltonian was written using the momentum representation of the polaron operators. 
Calculations in Ref. \cite{garwolaCompact2026} were focused on the eigenspectrum of light-matter models, and studying their ground state and thermal equilibrium properties, showing excellent performance of the CDH representation when focusing on both system's and cavity's observables. 

Complementing equilibrium properties, in the present study we focus on the dissipative dynamics captured by the CDH Hamiltonian. We carefully analyze the dynamical features captured by the method, its convergence and performance compared to numerically exact  and RC-model simulations, and insights gathered on the dynamics from the Liouvillian eigenspectrum.  

Consider a generic system, with the Hamiltonian $\hat H_S$, coupled to one or more bosonic baths,
\begin{equation}
    \hat H = \hat H_S + \sum_n\sum_k\left(\hat S_n t_{n,k}(\hat c_{n,k}^\dagger + \hat c_{n,k}) + \nu_k\hat c_{n,k}^\dagger\hat c_{n,k}\right).
\end{equation}
The different baths are labelled by $n$ and the modes of each bath by $k$. The bosonic creation (annihilation) operators of the baths are denoted by $\hat c_{n,k}^{\dagger}$  ($\hat c_{n,k}$)  with $\nu_k$ as the frequency of the $k$th mode. Displacements of the harmonic modes from bath $n$ are coupled to the system operator $\hat S_n$ with mode-dependent coupling strength,  $t_{n,k}$.

The system may be strongly coupled to the baths and we seek to study its dynamics.
To do so, the strong coupling is absorbed into the system by extracting a reaction-coordinate from each bath, defined such that $\lambda_n(\hat a_n^\dagger + \hat a_n) = \sum_kt_{n,k}(\hat c_{n,k}^\dagger + \hat c_{n,k})$.
The coupling strength and frequency of this mode are given by $\lambda_n$ and $\Omega_n$, respectively, which can be calculated from the spectral density $J_n(\omega)$ of the bath (for details see Ref. \cite{nazirReaction2018}).
The initial spectral density is taken to be a Brownian
\begin{equation}\label{eq:brownian}
    J(\omega) = \frac{4\gamma\lambda^2\Omega^2\omega}{(\omega^2-\Omega^2)^2+(2\pi\gamma\Omega\omega)^2},
\end{equation}
where $\gamma$ is the width parameter of the spectral density. 
However, it is important to note that the RC mapping can be operated on general spectral densities. Using the Brownian function, the interpretation of $\Omega$ and $\lambda$ is intuitive: They represent the central frequency of the bath and the system-bath coupling energy, respectively. After this transformation, the so-called RC Hamiltonian is given by
\bea
    \hat H_{\rm RC} &=& 
\hat H_S + \sum_n \left(\lambda_n\hat S_n(\hat a_n^\dagger +\hat a_n)\right) + \Omega_n\hat a_n^\dagger\hat a_n) 
        \nonumber \\ &+& \sum_n\tilde{\sum_k} \left(f_{n,k}(\hat a_n^\dagger+\hat a_n)(\hat b_{n,k}^\dagger+\hat b_{n,k})\right. \nonumber
        \\&+& \left.\omega_{n,k}\hat b_{n,k}^\dagger\hat b_{n,k}\right).
\eea
Here, $\hat b_{n,k}^{\dagger}$ is the bosonic creation operator for modes in the $n$-th residual bath,  while $a_n^{\dagger}$ are discrete bosonic modes for the RC associated with each bath.
The $\tilde\sum_k$ indicates a sum over the remaining non-RC modes of the bath.
The advantage of this mapping is that the coupling to the residual bath is made potentially weaker than in the original bath, thus allowing weak coupling methods such as the Redfield quantum master equation (QME) to simulate the dynamics.
For example, given an initial Brownian spectral density Eq. \eqref{eq:brownian}, the spectral density of the residual bath is Ohmic \cite{nazirReaction2018}
\begin{equation}
\label{eq:ohmic}
    J_\text{CDH}(\omega) = \gamma\omega e^{-\omega/\Lambda},
\end{equation}
where $\Lambda\rightarrow\infty$ is a large wavelength cutoff. Since the coupling energy in Eq. \eqref{eq:brownian} is given by $\lambda$, while the coupling strength in Eq. \eqref{eq:ohmic} is determined by (dimensionless) $\gamma$, that means a strongly coupled (large $\lambda$), but narrow (small $\gamma)$ bath transforms into a weakly-coupled residual bath.
In return, the system's Hilbert space is increased by incorporating RC levels in order for the dynamics to converge.
Notice that the RC Hamiltonian comprises two connected coupled systems: The system is coupled to the RC, and the RC is coupled to the residual bath.

The polaron transform allows to more effectively capture the dynamics of hybrid spin-boson objects \cite{xuNoncanonical2016}.
This transformation mixes the RC and the system, $\hat U = \exp\left[\sum_n\frac{\lambda_n}{\Omega_n}\hat S_n(\hat a_n^\dagger - \hat a_n)\right]$ once applied to the RC Hamiltonian.
The resulting Hamiltonian is subsequently truncated to $\bm{M}$ levels in the collective modes to generate the effective Hamiltonian
\begin{widetext}
\begin{multline}
    \hat H^{(\bm{M})}_\text{cdh} = (\hat H_S)^{(\bm{M})}_\text{cdh} + \sum_n\left(\Omega_n\left(\hat a_n^\dagger \hat a_n\right)^{(\bm{M})} - \frac{\lambda_n^2}{\Omega_n}(\hat S_n^2)^{(\bm{M})}_\text{cdh}\right) \\+ \sum_n\tilde{\sum_k}f_{n,k}\left(-\frac{2\lambda_n}{\Omega_n}(\hat S_n)^{(\bm{M})}_\text{cdh} + \left(\hat a_n^\dagger + \hat a_n\right)^{(\bm{M})}\right)(\hat b_{n,k}^\dagger + \hat b_{n,k}) + \sum_n\tilde{\sum_k}\omega_{n,k}\hat b_{n,k}^\dagger\hat b_{n,k}.
\end{multline}
\end{widetext}
Note that $\bm{M}$ lists the degree of truncation for each collective mode $n$.
Here, if $\hat O$ is an operator defined on the system + reaction coordinate's Hilbert space, then $(\hat O)_\text{cdh} = \hat U\hat O\hat U^\dagger$ is the operator transformed into the polaron frame.
Importantly, $(\hat H_S)_\text{cdh}$ now depends both on the system and on the collective mode.
In practice, one needs to choose the degree of truncation of the system Hamiltonian and coupling operators.
Since the polaron transformation is internal to the system + RC, the spectral density of the residual bath does not change, $J_\text{CDH}(\omega)=J_\text{RC}(\omega)$.
The blocks $(\hat O)_\text{cdh}^{(\bm{lr})}$---where \textbf{l} and \textbf{r} may be multi-indices labelling the coupled levels for each RC---are calculated directly by transforming the ladder operator into momentum space
\cite{garwolaCompact2026}, 
\begin{equation}
    (\hat O)^{(\bf{lr})}_\text{cdh} = \int \rd \bm{p}\braket{\bm{l}|\bm{p}}\hat U(\bm{p})\braket{\bm{p}|\hat O|\bm{p}}\hat U^\dagger(\bm{p})\braket{\bm{p}|\bm{r}}.
\end{equation}
Here, $\hat U(\bm{p}) = \exp(-i\sqrt{2}\sum_n\frac{\lambda_n}{\Omega_n}p_n\hat S_n)$ is the polaron unitary written in the momentum representation.
Concretely, one needs to evaluate the integrals either analytically or numerically to obtain the transformed operators,
\begin{multline}
    (\hat O)^{(\bf{lr})}_\text{cdh} = \\ \int\prod_n \rd p_nh^*_{l_n}(p_n)h_{r_n}(p_n)\hat U(\bm{p})\braket{\bm{p}|\hat O|\bm{p}}\hat U^\dagger(\bm{p})e^{-p_n^2}.
\end{multline}
where $h_n(p)$ is the degree $n$ Hermite polynomial.

In this work, we assume that the system's coupling operators $\hat S_n$ commute with each other. As such, the product $\hat U\hat O\hat U^\dagger$ can be decomposed into the action of each bath displacement operators acting on the eigenmodes of the coupling operators.
Restricting the calculation to a single bath, this gives
\begin{multline}\label{eq:cdh_block}
    (\hat O)^{(lr)}_\text{cdh} = \\\sum_{k,k'}\left(\ket{k}\bra{k}\hat O\ket{k'}\bra{k'}\right)\otimes\braket{l|\hat D\left(\frac{\lambda}{\Omega}(s_k-s_{k'})\right)|r},
\end{multline}
where $s_{k}$ and $\ket{k}$ are the eigenvalues and eigenvectors, respectively, of the coupling operator $\hat S$. The elements of the displacement operator are given by \cite{garwolaCompact2026}
\begin{equation}\label{eq:disp_elem}
    \braket{l|\hat D(\alpha)|r} = \begin{cases}
        \sqrt{\frac{r!}{l!}}\alpha^{l-r}L_r^{(l-r)}(|\alpha|^2)e^{-|\alpha|^2/2} & l \geq r, \\
        \sqrt{\frac{l!}{r!}}(-\alpha^*)^{r-l}L_l^{(r-l)}(|\alpha|^2)e^{-|\alpha|^2/2} & l < r,
    \end{cases}
\end{equation}
where $L_i^j(\alpha)$ are the generalized Laguerre polynomials.

In practice, Eqs. \eqref{eq:cdh_block} and \eqref{eq:disp_elem} are used with SymPy \cite{Sympy} to symbolically calculate matrix elements of the system's operator $(\hat O)_{\rm cdh}$. It is constructed to size $M$ as decided by the degree of truncation. Numerical values are substituted at the time of simulation. So far, we have described the transformation of operators that construct the Hamiltonian.

The dissipative dynamics of the CDH and RC models 
are simulated using the Redfield master equation solver in QuTiP \cite{qutip5} to provide the time evolution of the reduced density matrix, $(\rho(t))_{\rm cdh}$ and $(\rho(t))_{\rm RC}$.  
The bath is assumed to be thermal, prepared at temperature $T=1/\beta$, as we set the  Boltzmann constant $k_B=1$.
In RC simulations, the initial condition of the extended system, $\rho_{\rm RC}(0)$ is constructed by building a factorized system-RC state, with the RC thermally occupied
\bea \rho_{\rm RC}(0)= \rho_{S}(0)\otimes \rho^{(M)}_{\rm th}, \nonumber\\
\rho^{(M)}_{\rm th} = \frac{1}{Z_{\rm th}}
e^{-\beta \sum_{n=0}^{M-1} n\Omega |n\rangle \langle n|}.
\eea
Here, $Z_{\rm th}$ the corresponding partition function. This ensures that the RC mode population is normalized.
In CDH simulations, we construct 
\bea
\rho_{\rm cdh}(0)= \hat U'\left(\rho_S(0) \otimes \rho_\text{th}^{(M)}\right)\hat U'^\dagger,
\eea
with a truncated operator $\hat U'=\exp(\frac{\lambda}{\Omega}\hat S(\hat a^\dagger-\hat a)^{(M)})$,  calculated numerically.
From this initial conditions, the dynamics in the CDH representation evolves according to
\begin{equation} 
    (\dot\rho(t))_{\rm cdh} = \mathcal{L}_{\rm cdh}(\rho(t))_{\rm cdh},
\end{equation}
with $\mathcal{L}_{\rm cdh}$ the respective Liouvillian of the dynamics, including both coherent and dissipative terms generated in the cdh basis. A similar calculation generates the dynamics of RC models. This dynamics is compared to
numerically-exact dynamics calculated using the HEOM solver also in QuTiP \cite{lambert2023qutipbofin,qutip5}.

In the CDH representation, expectation values of system operators are evaluated from
\bea
\langle \hat A(t)\rangle&\equiv& {\rm Tr}[\hat A\rho(t)]
\nonumber\\
&=&
{\rm Tr}[(\hat U'\hat A \otimes \hat I_M\hat U'^{\dagger})
(\rho(t))_{\rm cdh}]
\nonumber\\
&=&
{\rm Tr}[(\hat A)_{\rm cdh}(\rho(t))_{\rm cdh}].
\eea
%
According to this definition, $(\hat A)_\text{cdh}$ is calculated by expanding its dimensionality to encompass both the system and the truncated model, then
applying the truncated polaron transformation 
$\hat U'$,  calculated numerically. This order of steps ensures that operators are transformed in a compatible way to the density matrix.
Finally, as mentioned above, the time evolution of the reduced density matrix is done with the CDH mapped Hamiltonian.
Throughout this work, we distinguish between the reaction coordinate and its physical interpretation as a cavity mode. In the general open-system
setting, the RC is a collective bosonic degree of freedom constructed from the
original harmonic bath to capture the dominant system-environment interaction. It therefore does not, in general, correspond to a single physical mode of the original environment. In light-matter settings, however, the RC Hamiltonian describes a system coupled to an individual cavity mode, which itself interacts with a residual environment and it is therefore lossy. In this work, we use the term ``RC" when discussing the general mapping, and we refer to the collective mode as a ``cavity mode" when emphasizing the physical light-matter interpretation. The CDH acronym denotes the compact dressed Hamiltonian representation in either pictures.

\section{Simulations}
\label{sec:Simulations}

When simulating dissipative dynamics, we use four methods:
\begin{enumerate}
    \item The hierarchical equations of motion method:  We use it as a numerically exact reference.
    \item RC method: We simulate the dynamics under the RC Hamiltonian with the Redfield quantum master equation. For larger systems at strong coupling, when HEOM simulations require excessive memory and time to converge,  converged RC results are used as a reference dynamics. We refer to these simulations by ``RC".
    \item CDH method: The Redfield QME is used to simulate the dynamics with effective Hamiltonians obtained from the CDH method. These results are identified by ``CDH" in the figures.
    \item Weak coupling limit: The Redfield QME is used on the original untransformed Hamiltonian (i.e., no CDH and no RC). Such simulations are denoted by ``Weak Coupling" to point out that we use a perturbative weak-coupling approach (though we purposely push parameters beyond this limit, to manifest the failure of this technique, thus the need to adopt nonperturbative methods).
\end{enumerate}

In terms of scaling of simulation with system size: if the original system has Hilbert space dimension $D$, then the vectorized  density matrix dimension is $D^2$ and the 
Redfield QME tensor is of dimension $D^2\times D^2$ (method 4).
When $M$ extra levels from the RC degree of freedom are included, the system's Hamiltonian dimension increases to $M D$, and consequently the Redfield tensor grows to $(M D)^2 \times (MD)^2$ (methods 2 and 3). However, CDH simulations typically require a value of $M$ smaller than what one needs to use for RC simulations.
For HEOM, the system dimension remains $D$. However,  using $L$ terms in the bath decomposition and an overall cutoff of $N$ in the hierarchy depth leads to a total of $\binom{N+L}{L}$ additional auxiliary density operators \cite{katoHierarchical2018, lambert2023qutipbofin} (method 1). Consequently, although these density matrices may be small, any increase in the amount of bath decomposition terms, or hierarchy depth cutoff, results in a large increase in the number of auxiliary operators needed.


\subsection{Spin-Boson Model}
\label{sec:TLS}

The spin-boson model, with a two-level system (spin) coupled to harmonic-oscillator baths, has applications in reaction rate chemistry, quantum optics, and quantum information processing \cite{nitzanChemical2006}.
Numerous techniques were developed to investigate the relaxation dynamics of the model, exploring different regimes, including the adiabatic and nonadiabatic limit, and the weak to strong coupling regimes. Techniques include quantum master equations of increasing complexity, such as the Redfield equation, the polaron-transformed Redfield equation \cite{xuNoncanonical2016,jangTheory2011} and the noninteracting blip approximation \cite{dekkerDynamics1987,boudjadaDissipative2014,montoya-castilloApproximate2016}. Beyond perturbative methods, numerically-exact techniques were developed and benchmarked by investigating the behavior of the spin boson model, including, e.g., iterative influence functional path integral techniques  
\cite{Makri20,strathearnEfficientNonMarkovianQuantum2018b,fuxTensor2023, makriTensor1995}, quantum Monte Carlo \cite{chenInchworm2017,goulkoTransient2025}, wavefunction methods \cite{wangCoherent2010,Wang17}, and the Hierarchical equation of motion (HEOM) approach \cite{Wu15,tanimuraNumerically2020,Gatto24,Thoss26}.

Here, we use the CDH mapping combined with Redfield QME to simulate the relaxation dynamics of the reduced density matrix, and compare to different methods, as described at the beginning of Sec. \ref{sec:Simulations}.

\subsubsection{Hamiltonian}

The spin-boson Hamiltonian is given by
\bea
\hat H = \Delta\hat\sigma_z + E\hat\sigma_x + \sum_k\left(\hat\sigma_xt_k(\hat c^\dagger_k+\hat c_k)+\nu_k\hat c_k^\dagger\hat c_k\right).
\label{eq:HSB}
\eea
The Hamiltonian of the system is $\hat H_S = \Delta\hat\sigma_z+E\hat\sigma_x$, and the operator coupling the system to the bath is $\hat S = \hat\sigma_x$. Extracting the RC mode gives
\begin{multline}\label{eq:spin-boson}
    \hat H_\text{RC} = \Delta\hat{\sigma}_z + E\hat{\sigma}_x + \Omega\hat a^\dagger\hat a + \lambda\hat\sigma_x\left(\hat a^\dagger+\hat a\right) \\+ \tilde{\sum_k}\left(f_k\left(\hat a^\dagger +\hat a\right)\left(\hat b_k^\dagger+\hat b_k\right) + \omega_k\hat b^\dagger_k\hat b_k\right).
\end{multline}
In this representation, the system Hamiltonian is given by the first four terms: the original spin and its internal tunneling term, the added collective mode, and the coupling of this mode to the spin,
\begin{equation}\label{eq:HS_RC}
    \hat H^S_\text{RC} = \begin{pmatrix}
        \Delta\hat\sigma_z & \lambda\hat\sigma^x & 0 &\\
        \lambda\hat\sigma^x & \Delta\hat\sigma_z & \sqrt{2}\lambda\hat\sigma^x & \cdots\\
        0 & \sqrt{2}\lambda\hat\sigma^x & \Delta\hat\sigma_z & \\
        & \vdots & &\ddots
    \end{pmatrix} + E\hat\sigma_x+\Omega\hat a^\dagger\hat a.
\end{equation}
The next step in creating the CDH is applying the polaron transformation,  $\hat U = \exp(\frac{\lambda}{\Omega}(\hat a^\dagger - \hat a)\hat\sigma_x)$ leading to the CDH representation, 
\begin{multline}\label{eq:spin_boson_cdh}
    \hat H_\text{cdh} = \Delta(\hat\sigma_z)_\text{cdh} + E\hat\sigma_x + \Omega\hat a^\dagger\hat a 
    \\+ \tilde{\sum_k} f_k\left(\hat a^\dagger + \hat a - \frac{2\lambda}{\Omega}\hat\sigma_x\right)(\hat b_k^\dagger + \hat b_k) + \tilde{\sum_k}\omega_k\hat b_k^\dagger\hat b_k.
\end{multline}
Since $\hat \sigma_x$ and $\hat U$ commute, only $\hat \sigma_z$ is transformed.
Explicitly, the polaron-transformed system Hamiltonian (first three terms in Eq. (\ref{eq:spin_boson_cdh})) is given by
\begin{widetext}
\begin{equation}\label{eq:HS_CDH}
    \hat H^S_\text{cdh} = \Delta e^{-2\epsilon^2}\begin{pmatrix}
        \hat\sigma_z & 2\epsilon i\hat\sigma_y & 2\sqrt 2\epsilon^2\hat\sigma_z & \ldots \\
        -2\epsilon i\hat\sigma_y & (1-4\epsilon^2)\hat\sigma_z & -2\sqrt2(2\epsilon^2-1)i\hat\sigma_y & \ldots \\
        2\sqrt2\epsilon^2\hat\sigma_z & 2\sqrt2(2\epsilon^2-1)i\hat\sigma_y & (8\epsilon^4-8\epsilon^2+1)\hat\sigma_z & \ldots \\
        \vdots & \vdots & \vdots & \ddots
    \end{pmatrix} + E\hat\sigma_x+\Omega\hat a^\dagger\hat a,
\end{equation}
\end{widetext}
where $\epsilon=\lambda/\Omega$.
Notice the presence of additional off-diagonal terms, beyond those present in Eq. \eqref{eq:HS_RC}, which was tridiagonal due to the coupling to $\hat a+\hat a^\dagger$.
These additional off-diagonal terms capture higher (e.g., multi-photon) transitions, with even-number transitions coupled via $\hat\sigma_z$ and odd-number transitions by $\hat\sigma_y$.
Another outcome of strong coupling is that the bare spin splitting $\Delta$ is rescaled by the factor $\exp(-2\epsilon^2)$. 

Returning to Equation (\ref{eq:spin_boson_cdh}), the system-bath coupling operator in the polaron frame is given by
\begin{equation}
    \hat S_\text{cdh} = \begin{pmatrix}
        -\frac{2\lambda}{\Omega}\hat\sigma_x & 1 & 0 & \ldots \\
        1 & -\frac{2\lambda}{\Omega}\hat\sigma_x & \sqrt{2} & \ldots \\
        0 & \sqrt{2} & -\frac{2\lambda}{\Omega}\hat\sigma_x & \ldots \\
        \vdots & \vdots & \vdots & \ddots
    \end{pmatrix}.
\end{equation}
Besides the RC displacement operator that couples to the residual bath,  the polaron transformation introduces a direct coupling term between the spin and the environment, $-\frac{2\lambda}{\Omega}\hat\sigma_x(\hat b_k^{\dagger}+\hat b_k)$. 
Taken together, the CDH Hamiltonian captures the rescaling of the system, hybridization of the spin and RC mode, and relaxation to the bath --- at a strength weaker than in the original picture.

As mentioned in Section \ref{sec:mapp}, for the original spectral density, the Brownian function, Eq. \eqref{eq:brownian} is adopted. 
After the RC transformation, the spectral density of the residual bath is Ohmic, given by Eq. \eqref{eq:ohmic}.
This spectral density is independent of the RC frequency and the system-reaction coordinate coupling strength parameter, $\lambda$.
Thus, for sufficiently small $\gamma$ 
the system remains in the weak coupling regime with respect to the residual bath, regardless of the coupling in the original system.

In simulations, we typically consider a spin with energy splitting $\Delta=1$, temperature $T=1/\beta=\Delta$, RC frequency $\Omega=2$--$10$, and bath width parameter $\gamma=0.05$. These parameters place the system within the regime of validity of the Redfield equation in the CDH representation: the residual system-bath coupling is perturbatively weak, $\gamma\Delta\ll1$. 
The additional Markov approximation requires the residual bath correlation time to be short compared with the relevant relaxation timescales of the dressed system. The accuracy of these approximations is assessed by comparisons with numerically exact HEOM calculations. 

\begin{figure*}[ht]
    \centering
\includegraphics[width=0.95\linewidth]{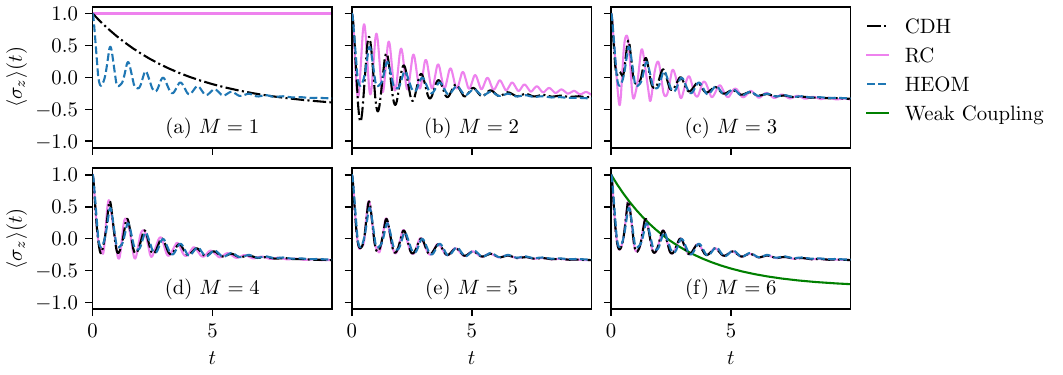}
    \caption{Spin-boson simulations demonstrating rapid  convergence of the CDH representation for $E=0$, $\lambda=5\Delta$, $\Omega=9\Delta$, $\gamma=0.05$, and $\beta = 1/\Delta$. (a)-(f) The CDH (dashed-dotted black lines) and RC (solid purple line) are executed with truncations $M=1-6$, and compared to HEOM simulations (dashed blue line). 
    As $M$ is increased, both representations converge toward the same result, since the underlying untruncated Hamiltonians are related by a unitary transformation, yet the CDH solution converges faster. This is particularly noticeable for $M=2$ and $M=3$. (f) The weak coupling Redfield QME solution is also included (full green line). It shows a poor match with exact results, indicating that the dynamics is well into the strong coupling regime.}
    \label{fig:heom_match}
\end{figure*}

\subsubsection{Relaxation Dynamics: Off Resonance Situations}

Since the CDH Hamiltonian is obtained from a unitary transform on the RC Hamiltonian, they both converge to the same solution when sufficient levels are included. 
However, as we show next, the CDH Hamiltonian allows faster convergence to that limit.

In Fig. \ref{fig:heom_match} we plot the polarization dynamics 
$\langle \sigma_z(t) \rangle\equiv {\rm Tr}[\hat \sigma_z\rho(t)]=
{\rm Tr}[(\hat{\sigma}_z)_{\rm cdh}(\rho(t))_{\rm cdh}]$,
with $(\rho(t))_{\rm cdh}$ the solution of the Redfield equation for the system's evolution using the CDH framework. As discussed in Section \ref{sec:mapp}, the truncated polaron transformation operator $\hat U'=\exp(\epsilon\hat\sigma_x(\hat a^\dagger-\hat a)^{(M)})$ is used to ensure a unitary transformation of operators into the polaron frame.
Results presented are  derived from the CDH (dashed-dotted black) and RC (pink) representations with increasing $M$, the degree of truncation. These results are compared to numerically exact simulations produced by the hierarchical equations of motion method (dashed blue) \cite{tanimuraNumerically2020} and to the weak coupling limit in panel (f) (full green).

The CDH solution with only $M=2$ levels is already remarkably close to the HEOM solution, except for a larger amplitude at short times and a slight phase discrepancy. 
In contrast, RC simulations when performed at $M=1-4$ noticeably deviate from the exact dynamics in magnitude and phase shift.

\begin{figure*}[ht]
    \centering
\includegraphics[width=0.95\linewidth]{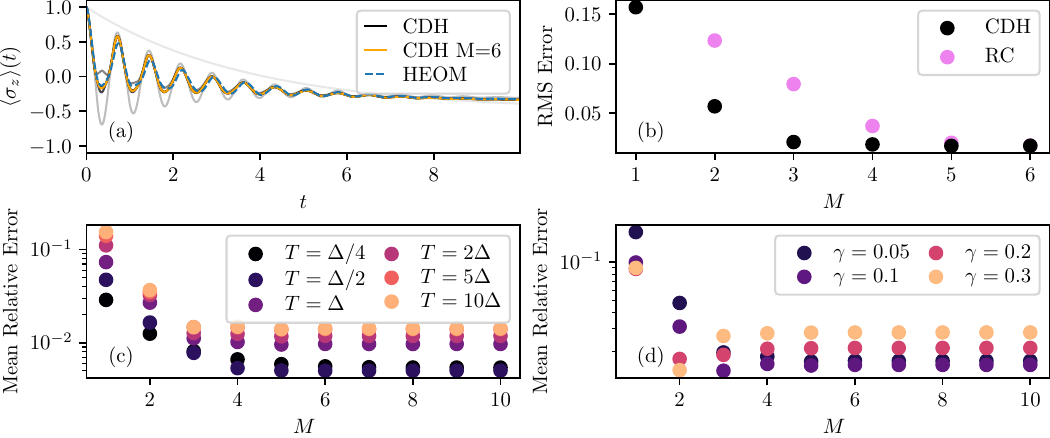}
\caption{Convergence analysis of spin-boson simulation results.
(a) Dynamics of the CDH method (grey lines) approaching the HEOM solution with the $M=6$ CDH result highlighted in yellow.
(b) Comparison between the RMS error (Eq. (\ref{eq:rms_error})) for the RC and CDH methods for different $M$ values. Simulations in panels (a) and (b) are the same as those shown in Fig. \ref{fig:heom_match}.
(c) Mean relative error (Eq. (\ref{eq:mre_error})) of CDH with respect to the HEOM, plotted for each $M$ for varying values of $T$. 
(d) Mean relative error (Eq. (\ref{eq:mre_error})) of CDH relative to HEOM plotted for each $M$ for varying values of $\gamma$.
We use $E=0$, $\lambda=5\Delta$ and $\Omega=9\Delta$, and a time step $\Delta t=1/100\Delta$ in all simulations, with $\gamma=0.05$ and $\beta = 1/\Delta$ in (a) and (b), and varying $T$ and $\gamma$ in (c)-(d). The temperature simulations are run until time $t=10/\Delta$ while the varying $\gamma$ simulation are run until time $t=6/\Delta$. Corresponding dynamics  are shown in Appendix \ref{s:T_and_g}.
}
    \label{fig:meth_temp_gamma}
\end{figure*}

In Fig. \ref{fig:meth_temp_gamma}, the data from Fig. \ref{fig:heom_match} are combined and analyzed. First, in panel (a) $M=1-5$ degree CDH results are plotted in light shades of grey, while $M=6$ results are highlighted in yellow. HEOM simulations are also included.
As mentioned above, even at the second level, the CDH solution captures to a good approximation the oscillation frequency, relaxation rate, and steady-state value in agreement with the exact solution. 
The error is quantified in panel (b) using the root-mean-square error (RMS) compared to HEOM simulations, plotted against the number of levels $M$ in the truncation, 
\begin{equation}\label{eq:rms_error}
    \text{RMS Error} = \sqrt{\frac{\sum_{i=1}^N(\braket{\sigma_z}(t_i) - \braket{\sigma_z}_\text{HEOM}(t_i))^2}{N}}.
\end{equation}
Here, $N$ is the total number of time steps, $t_i$. The deviation of the two methods (CDH and RC) from HEOM is similar by $M=6$, which is expected given the energy parameters chosen for the system and the temperature of the bath.  For lower truncations, in particular $M=2$ and 3, the CDH method has a smaller error than the RC, as shown in Fig. \ref{fig:heom_match}. 
Although CDH can be made more accurate with increasing $M$, the application of Redfield QME amounts to a Markovian and weak coupling assumption with respect to the residual bath, which explains the error floor.
In principle, one could go beyond these approximations, by using a higher level method to study the dissipative dynamics. Here, we contain ourselves with the Redfield QME and assess the error associated with the weak coupling Markovian approximation in Fig. \ref{fig:heom_match} (c)-(d). The Mean Relative Error of the spin's population $P = \frac{\braket{\hat\sigma_z}+1}{2}$ is used as a measure of error,
\begin{equation}\label{eq:mre_error}
    \text{Mean Relative Error} = \frac{1}{N}\sum_i\frac{|P(t_i)-P_\text{HEOM}(t_i)|}{P_\text{HEOM}(t_i)} 
\end{equation}
This relative measure is used, rather than an absolute measure, since the numerically exact solution is different for each set of parameters, as one varies the temperature and the spectral width of the bath. Using the population further avoids the issue of the relative error blowing up for equilibrium values near zero (corresponding to a spin population of 1/2). For this comparison, only the magnitude of the deviation is considered, not the relative direction of the error (over/undershoot), as such the absolute value of the relative error is averaged in Eq. (\ref{eq:mre_error}). The exact dynamics from which the error was calculated is shown in Appendix \ref{s:T_and_g}. 

We find that the CDH method performs best for $T = \Delta/2$ and worsens for both lower and higher temperatures. Since the weak coupling Redfield master equation is being used to simulate the CDH dynamics, it would be expected to perform better at high $T$, where the bath relaxes quickly and dynamics should be overall more Markovian, compared to low $T$. Understanding why the error increases with temperature remains to be investigated.

For small $M$, the method performs better for larger $\gamma$, but converged solutions perform better for smaller $\gamma$, as expected. Note that the error saturates as the RC and CDH methods agree. More studies with  less numerically-sensitive error measures would be necessary to determine and understand in detail the performance of the CDH framework with respect to $T$ and $\gamma$.

\begin{figure*}[ht]
    \centering
\includegraphics[width=0.95\linewidth]{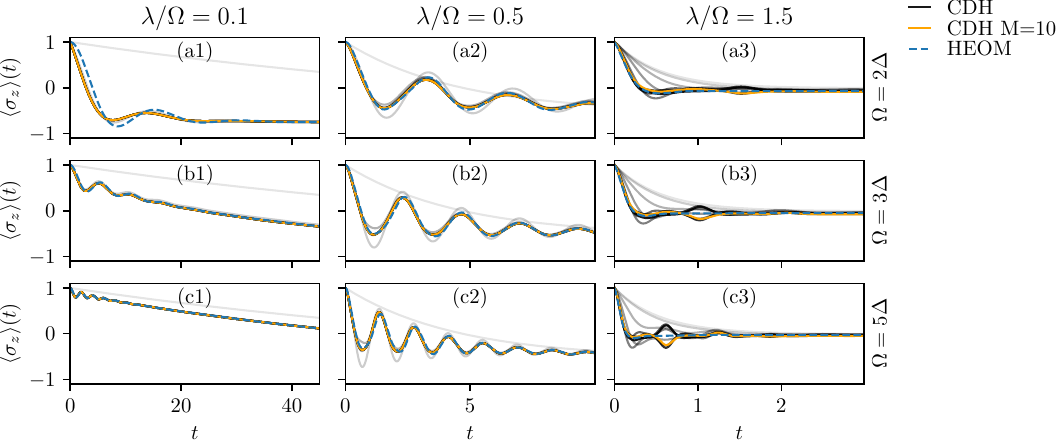}
    \caption{
Systematic analysis of the CDH performance with changing bath frequency and coupling strength. CDH results with $M=1-10$ are shown (black, $M=10$ in orange) for different ratios of $\lambda/\Omega$. From left to right $\lambda/\Omega = 0.1$, 0.5, 1.5, and from top to bottom $\Omega = 2\Delta$, $3\Delta$, $5\Delta$. We use  $E=0$, $\gamma=0.05$, and $\beta=1/\Delta$ in all panels. The benchmark is given by the HEOM (dashed blue). Two trends are evident: the convergence becomes slower with increasing coupling strength and when approaching resonance.
Note that different columns use different time ranges.
}
    \label{fig:conv}
\end{figure*}

\begin{figure*}[ht]
    \centering
\includegraphics[width=0.95\linewidth]{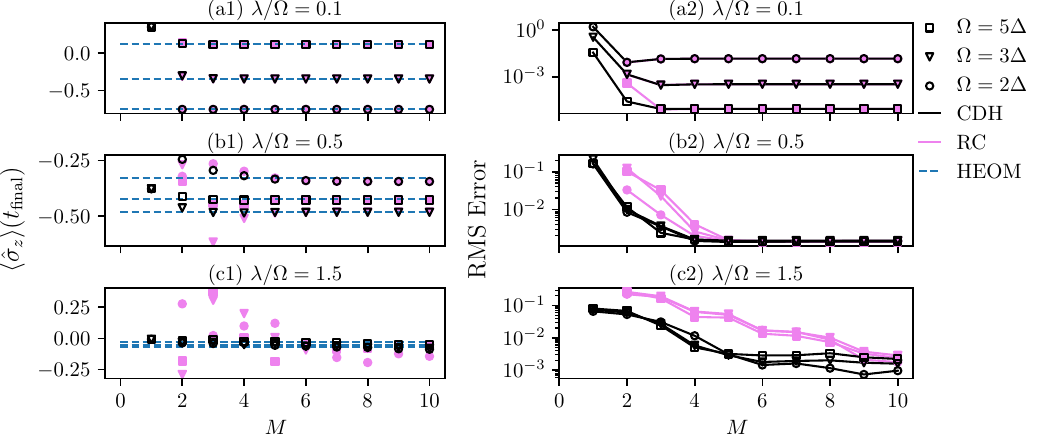}
\caption{RC and CDH performance compared to numerically exact results, for simulations shown in Fig. \ref{fig:conv}. 
(a1)-(c1) Final-time values of each simulation, approximating steady state values, are plotted vs. $M$. At strong coupling, the CDH produces results closer to the correct steady-state than RC. 
(a2)-(c2) Root mean square error plotted against $M$. The CDH and RC require the same number of levels to reach convergence, but generally, the CDH method has a smaller error leading up to convergence.} 
    \label{fig:cumul_alt}
\end{figure*}
\subsubsection{Relaxation Dynamics: Near Resonance Scenarios}

The performance of the CDH framework is systematically analyzed in 
Figs. \ref{fig:conv} and \ref{fig:cumul_alt} considering various coupling strengths and bath-frequency regimes. For the characteristic frequency of the bath $\Omega$, we test cases from resonance $\Omega=2\Delta$, close-to-resonance $3\Delta$, to off-resonance scenarios $5\Delta$. The system-bath coupling strength is modified from weak to strong with $\lambda/\Omega = 0.1$, 0.5, and 1.5.

Although the strong coupling regime is defined as $\lambda/\Delta \gg 1$ \cite{friskkockumUltrastrong2019, forn-diazUltrastrong2019}, the polaron transformation rescales the coupling strength by $\Omega$, as seen in the off-diagonal terms of Eq. \eqref{eq:HS_CDH} and the reintroduced system-residual bath coupling in Eq. \eqref{eq:spin_boson_cdh}. Thus, when talking about strong coupling in the CDH context, it is practically useful to focus on the ratio $\epsilon \equiv \lambda/\Omega$ as a metric.

We present results with the CDH method from $M=1$ and up to $M=10$ and compare them with the RC and HEOM simulations. 
Two main trends are identified in Fig. \ref{fig:conv}: The CDH method requires more resources (high $M$) at strong coupling both close to resonance (a3) and for off-resonance situations (c3).
In particular, for $\lambda/\Omega=0.1$ but at resonance with $\Omega=2\Delta$ (a1),
there are short time oscillations in the converged CDH $M=10$ solution that are not present in the HEOM and that, despite being weak coupling, show worse early time dynamics than away from resonance, see (b1) and (c1). 

In Fig. \ref{fig:cumul_alt}(a1)-(c1) we display the final point of the CDH and RC simulations compared to HEOM. This allows us to compare the long time  behavior of these methods, representing the steady state. At strong coupling, the CDH converges more quickly to the correct steady-state than the RC does, though at weaker coupling the difference is minimal.
The challenge of convergent simulations close to resonance is observed in  Fig. \ref{fig:cumul_alt} (a2)-(c2), showing the RMS error as a function of $M$. Across the different values of $\Omega$ and for different coupling strengths, the CDH method consistently has a lower error than RC. We also find that at weak coupling the performance is worse at resonance compared to the off-resonance case, see Fig. \ref{fig:cumul_alt} (a2), and that convergence requires higher $M$ as the coupling strength is made stronger.

\begin{figure*}[ht]
    \centering
\includegraphics[width=0.95\linewidth]{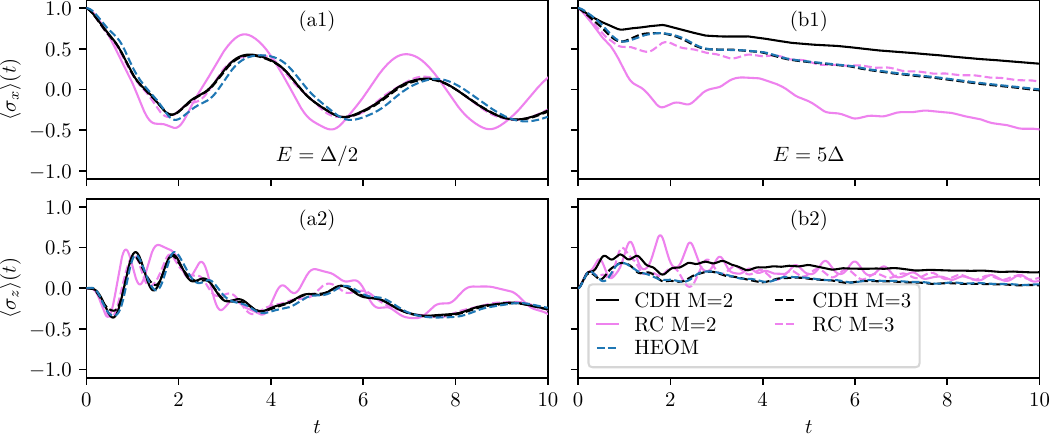}
    \caption{Dynamics of the biased spin boson model with a comparison between the CDH, RC, and HEOM methods for both (a1)-(b1) coherence $\langle \sigma_x(t)\rangle $ and  (a2)-(b2) polarization, $\langle \sigma_z(t)\rangle $. Parameters are $\Omega=7\Delta$, $\lambda=3\Delta$, $\gamma=0.05$, and $\beta=1/\Delta$, with (a1)-(a2) $E=\Delta/2$ and (b1)-(b2) $E=5\Delta$. The CDH and RC are simulated with $M=2$ and 3 levels.
    }
    \label{fig:meth_w_bias}
\end{figure*}

\subsubsection{Relaxation Dynamics with Non-Zero Bias}

So far, we focused on the unbiased spin-boson model with $E=0$ in Eq. (\ref{eq:HSB}).
Complementing this, we exemplify the behavior of the spin-boson model with non-zero bias in Fig. \ref{fig:meth_w_bias} using $E=\Delta/2$ (left) and $E=5\Delta$ (right). 
Here, we study both the population dynamics, $\langle \sigma_z(t) \rangle $, and the coherence dynamics, $\langle \sigma_x(t) \rangle $. As an initial condition, the spin is prepared in the $+x$ eigenstate. We focus only on small $M$ values and show that the $M=3$ CDH calculation is more accurate than the corresponding RC calculation. 
In general, already with a small $M$ value, the CDH method is capable of correctly capturing the magnitude, oscillation frequency, and population and decoherence rates compared to numerically exact simulations. 


\subsubsection{Liouvillian Spectrum and Relaxation Rate Constant}

The advantage of the CDH method over the RC approach becomes increasingly pronounced as the coupling strength grows, reflecting the ability of the polaron-transformed hybridized basis to capture strong-coupling effects more effectively.
Another central outcome of the method is the estimation of rate constants. The eigenvalue spectrum of the Liouvillian superoperator provides the relaxation timescales and the corresponding dynamical modes. Working in the Hamiltonian energy basis, we propose here to use the eigenvalue with the smallest real part that affects the population, which corresponds to the slowest dissipation process, as a proxy for the reaction rate constant.

The definition and extraction of reaction rates from quantum dynamics is nontrivial, particularly when population dynamics exhibit coherent oscillations rather than simple exponential relaxation. Conventional approaches include the extraction of the rate from population kinetics \cite{jangTheory2008,lawrenceCalculation2019,lindoyQuantum2023} 
and formally exact formulations based on flux correlation functions  \cite{millerQuantum1983,wangCalculation2006,keNonequilibrium2022}. 
Within the Markovian embedding framework employed here, the Liouvillian spectrum provides an alternative feasible characterization of the intrinsic dynamical timescales: population relaxation rates can be associated directly with the decay eigenmodes that overlap the relevant population observable, without fitting the time-domain dynamics.

\begin{figure*}[ht]
    \centering
\includegraphics[width=0.95\linewidth]{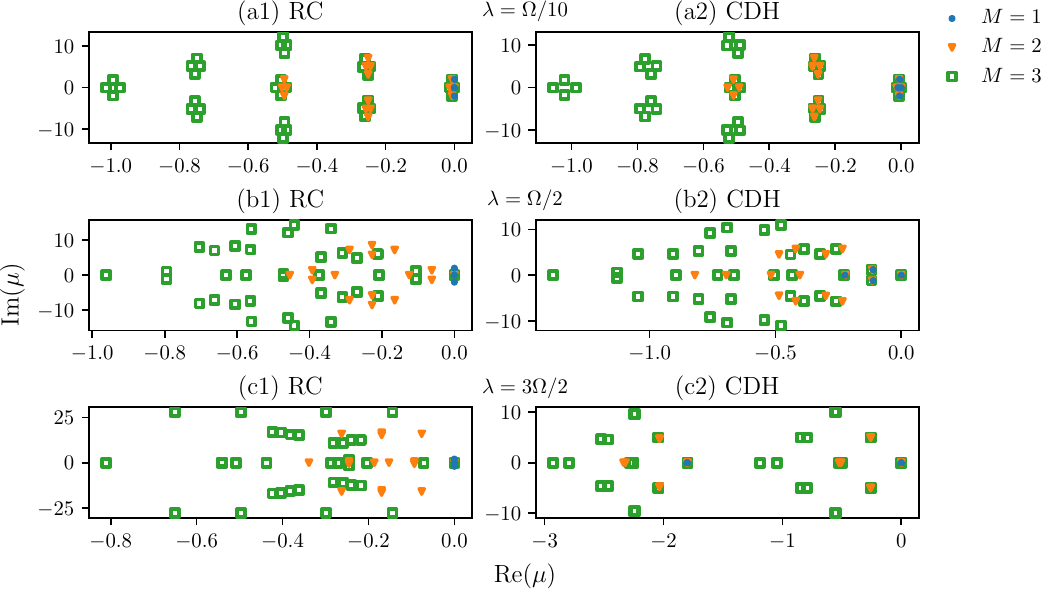}
    \caption{Eigenvalue spectrum of the RC (left) and CDH (right) Redfield superoperators (Liouvillian) of the spin-boson model at $M=1$, 2, and 3, with $\Omega=5\Delta$, $\beta=1/\Delta$, $\gamma=0.05$, and $E=0$. 
    (a) $\lambda=\Omega/10$, (b) $\lambda=\Omega/2$, (c) $\lambda=3\Omega/2$, corresponding to panels (c1)-(c3) of Fig. \ref{fig:conv}. For each $M$ there are $4M^2$ eigenvalues.
    }
    \label{fig:red_spec}
\end{figure*}

Using the vectorized representation of the density operator, the eigenvalues $\mu$ and the right eigenvectors $v$ of the Liouvillian are given by
\bea
\mathcal{L} {\bf v}= \mu {\bf v}.
\eea
The real part of the eigenvalues $\mu$ corresponds to the negative of the relaxation rates, while the imaginary part provides the oscillation frequencies,
\bea
\rho(t)- \rho_{\rm ss} = \sum_{j\neq 0} c_j e^{\mu_j t}
\eea
Here, $\rho_{\rm ss}$ is the steady state solution.
The coefficients $c_j$ depend on the initial preparation.

Figure \ref{fig:red_spec} shows the low-lying eigenvalues, which govern the long-time approach to the steady state, computed using the Redfield superoperator. We find that these eigenvalues are remarkably stable within the CDH method even for small values of $M$, moving only slightly as $M$ increases. In contrast, the RC approach requires convergence with respect to $M$ before it can accurately reproduce even the slowest relaxation modes.

\begin{figure*}[ht]
    \centering
    \includegraphics[width=0.95\linewidth]{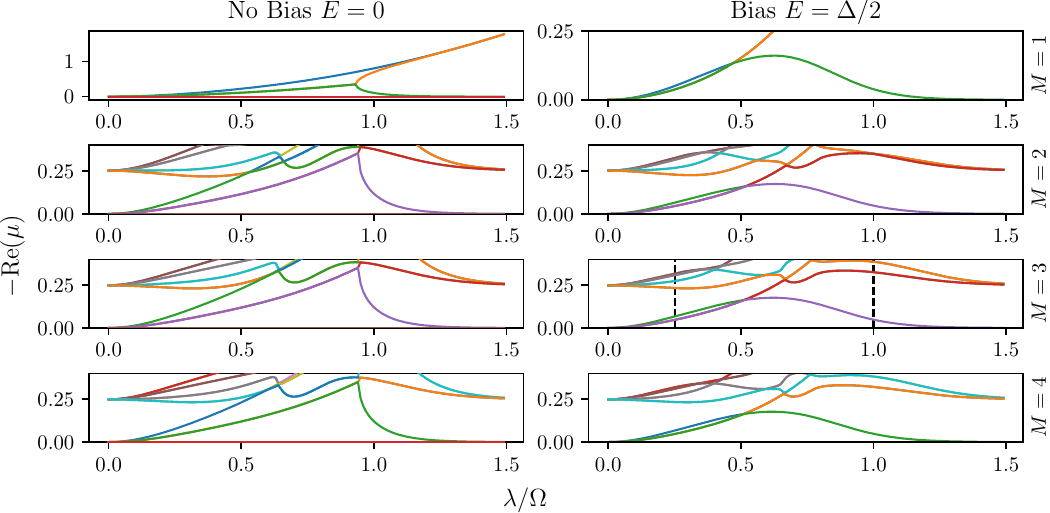}
    \caption{Numerically calculated negative real parts of the eigenvalues $\mu_i$ of the Redfield superoperator for the spin boson model with $\Omega=5\Delta$, $\gamma=0.05$, $\beta=1/\Delta$ and $E=0\Delta$ and $\Delta/2$ plotted against $\lambda/\Omega$ for $M=1$, 2, 3, and 4. The eigenvalues are coloured in order of the magnitude of their real part such that two points with the same colour may not correspond to the same analytic eigenvalue. The black dashed lines indicate the parameters for which the eigenmodes are shown in Fig. \ref{fig:eig_vecs}.}
    \label{fig:eig_v_m_bias}
\end{figure*}

The parameters used correspond to those presented in Figs. \ref{fig:conv} (c1)-(c3), capturing weak to strong coupling.
Since the dimensionality of the Hamiltonian is $2M\times2M$, the superoperators are size $4M^2\times4M^2$. Therefore, there are $4M^2$ eigenvalues at each choice of $M$, some seem overlapping. At the strongest coupling, eigenvalues come in nearly identical pairs, which overlap in Fig. \ref{fig:red_spec}. This can be seen in Fig. \ref{fig:eig_v_m_bias}, where the real parts of different eigenvalues approach each other at strong coupling. 


In weak coupling regime, see Fig. \ref{fig:red_spec}(a1)-(a2), the eigenvalue spectra generated  by both the CDH and RC methods are quite robust with $M$ when it is increased from 1 to 3. Note that increasing $M$ adds more negative (higher frequency) eigenvalues to the spectrum.
In the opposite strong coupling regime, $\lambda=3\Omega/2$, the CDH spectrum remains well organized with increasing $M$.
For example, the imaginary part of the CDH spectrum is separated into bands at $0$, $\pm\Omega$ and $\pm2\Omega$, and splits into separate groups of slow decaying and faster decaying modes as one moves to a more negative real part of $\mu$. In contrast, the RC spectrum for $M=1,2,3$ misses these features and changes significantly with $M$, including major changes to low-lying eigenvalues. As such, within the RC method, only when $M$ is large enough can one use the smallest nonzero magnitude $\mu$ to represent the rate. In contrast, the rate could be extracted efficiently (and even analytically) from CDH simulations with as small as $M=1,2$.


\begin{figure*}[ht]
    \centering
    \includegraphics[width=0.95\linewidth]{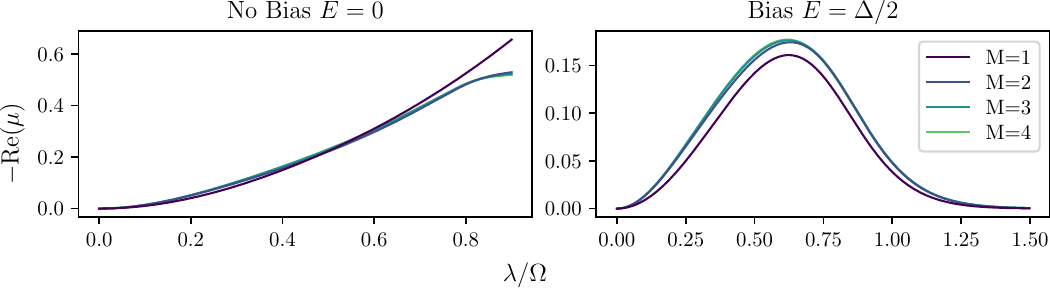}
    \caption{Extraction of population relaxation rate constant as the smallest non-zero real eigenvalue. We consider cases with and without bias as $\lambda$ varies, extracted from Fig. \ref{fig:eig_v_m_bias}. Other parameters are $\Omega =5\Delta$, $\gamma=0.05$, $\beta=1/\Delta$. When $E=0$, at larger couplings than plotted, the eigenvalues of the first coherence eigenmodes become real, and the smallest non-zero real eigenvalue no longer corresponds to the population relaxation rate.}
    \label{fig:mu_2}
\end{figure*}

The fast  convergence of the smallest eigenvalues, already for small values of $M$, highlights that the CDH method is suitable to extract relaxation rates. In Fig. \ref{fig:eig_v_m_bias} the negative real parts of the eigenvalues of the Redfield superoperator using the CDH method are calculated numerically and plotted as a function of $\lambda/\Omega$ for $M=1-4$ for the spin-boson model with and without an energy bias. In Fig. \ref{fig:mu_2} the smallest (in magnitude) non-zero real eigenvalue is extracted from these simulations. For the biased system, this eigenvalue is associated with the relaxation rate of the system, and the corresponding eigenmode is the longest lived. The determination of which eigenvalues correspond to populations or coherences was made by looking at the eigenmodes in the energy basis of the Hamiltonian $\hat H^S_\text{cdh}$. Eigenmodes corresponding to populations are diagonal in the energy basis.
To complement the eigenvalues, the eigenmatrices of the Redfield tensor are presented for $\lambda/\Omega=0.25$ and 1 in Appendix \ref{s:eigenmats}. 

In the unbiased case,
the eigenvalues are more crowded, and we identify the rate only up to a particular coupling strength, $\lambda^*$, above which the eigenvalues of a pair of coherence eigenmodes become real.
In Appendix \ref{s:eigenvalues} we discuss this point in more detail and obtain $\lambda^*\approx4.66$ ($\lambda^*/\Omega\approx0.932$) for $M=1$ for the parameters used in Figs. \ref{fig:eig_v_m_bias} and \ref{fig:mu_2}. 


\subsection{Multiple Spins}
\label{sec:multspin}


%

Having established the feasibility and accuracy of the CDH method for the  spin-boson model, we now turn to systems comprising multiple interacting spins. In this setting, coupling to a common bosonic environment can generate effective interactions between otherwise independent degrees of freedom, giving rise to collective dynamical phenomena that are absent in the single-spin problem. The CDH representation is particularly useful for studying such effects, as the dressing transformation incorporates the strongly coupled environmental modes directly into the effective system Hamiltonian, making bath-mediated interactions explicit in the CDH, while retaining a compact description of the dissipative dynamics.

We begin with a system of two spins that interact directly (coherently) with each other while also coupled to a common bosonic environment. Many-spin generalizations of this setup are closely related to Ising- or Heisenberg-Dicke models, depending on the form of the direct spin-spin interaction. In the absence of direct interactions, the corresponding many-spin model reduces to the Dicke model and its variants, including disordered realizations.

\begin{figure*}[ht]
    \centering
\includegraphics[width=0.95\linewidth]{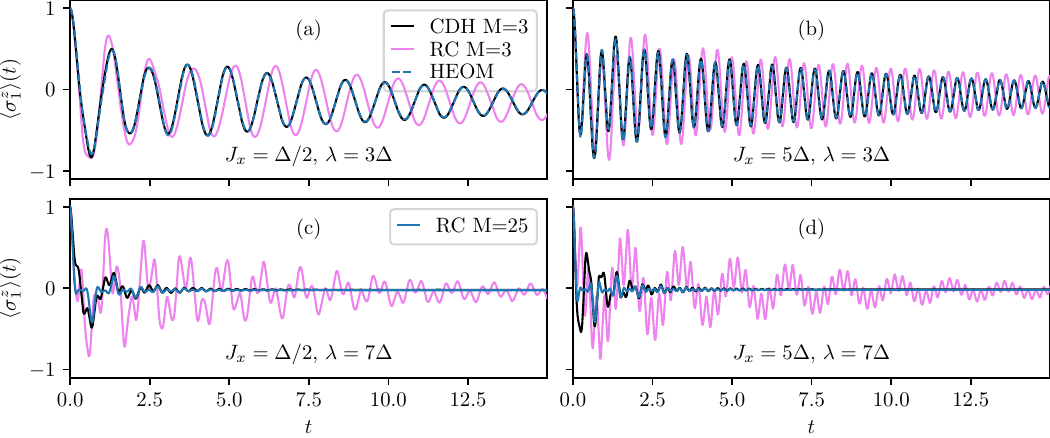}
\caption{Comparison of the CDH and RC methods for two spins coupled to a common bosonic bath and interacting through a nearest-neighbor $J_x\hat{\sigma}^x_1\hat{\sigma}_2^x$ coupling. We set $\Delta_2=\Delta_1\equiv\Delta$, $\Omega=9\Delta$, $\gamma=0.05$, and $\beta=1/\Delta$. The coupling parameters are 
(a) $\lambda=3\Delta$, $J_x=\Delta/2$; 
(b) $\lambda=3\Delta$, $J_x=5\Delta$; 
(c) $\lambda=7\Delta$, $J_x=\Delta/2$; and 
(d) $\lambda=7\Delta$, $J_x=5\Delta$. 
In panels (c) and (d), converged RC dynamics with $M=25$ are used as the reference since HEOM calculations did not converge. 
Across all four cases, the CDH method approaches the reference calculation with a smaller truncation than the RC method.   }
    \label{fig:2_spins}
\end{figure*}
\subsubsection{Synchronization of Two Spins}

We begin with the minimal two-spin setting, which already displays nontrivial collective dynamics induced by the common environment.
As demonstrated in Ref.~\cite{brenesBathinduced2024}, coupling to a common environment generates an effective bath-mediated interaction between the spins. This induced interaction  leads to synchronization, even in the absence of a direct coherent coupling between them. This dynamical synchronization is manifested by a common oscillation frequency in the time-dependent magnetization of the two spins.

This synchronization dynamics was previously investigated in e.g., Ref.~\cite{brenesBathinduced2024} using the effective-Hamiltonian approach with restrictive truncation, $M=1$. At weak system-bath coupling, this approximation accurately captured the two-spin dynamics, including the emergence of synchronization. However, in strong coupling, while the description $M=1$ reproduced the steady state and the synchronization frequency, it did not capture the transient dissipative dynamics and missed prominent dynamical features. Here, we revisit this problem using the CDH framework with $M>1$, and show that we can now systematically recover dynamical features at strong coupling, and assess the convergence of the synchronization dynamics with the size $M$ of the dressed representation.

We consider the following Hamiltonian.
\begin{multline}\label{eq:2_spin_ham}
    \hat H = \Delta_1\hat\sigma_1^z + \Delta_2\hat\sigma_2^z - J_x\hat\sigma_1^x\hat\sigma_2^x \\+ (\hat\sigma_1^x+\hat\sigma_2^x)\sum_kt_k(\hat c_k^\dagger+\hat c_k) + \sum_k\nu_k\hat c_k^\dagger\hat c_k.
\end{multline}
After applying the CDH transformation with $\hat U = \exp(\frac{\lambda}{\Omega}(\hat\sigma^x_1+\hat\sigma^x_2)(\hat a^\dagger - \hat a))$, the Hamiltonian of the system is 
\begin{multline}\label{eq:2_spins_trans}
    \hat H^S_\text{cdh} = \Delta_1(\hat\sigma_1^z)_\text{cdh} + \Delta_2(\hat\sigma_2^z)_\text{cdh} \\- \left(J_x+\frac{2\lambda^2}{\Omega}\right)\hat\sigma_1^x\hat\sigma_2^x + \Omega\hat a^\dagger\hat a.
\end{multline}
The system operator that couples to displacements in the residual bath is given by the two-spin generalization of Eq. (\ref{eq:spin-boson}),
\begin{equation}
    \hat S_\text{cdh} = -\frac{2\lambda}{\Omega}(\hat\sigma_1^x + \hat\sigma_2^x)+\hat a^\dagger + \hat a.
\end{equation}
Notice that after the transformation, the $XX$ coupling term is increased to 
$J_{x} + \frac{2\lambda^2}{\Omega}$, reflecting the bath mediated coupling between spins, favoring a ferromagnetic order at asymptotically strong coupling along this direction.


In Fig.~\ref{fig:2_spins}, we examine the population dynamics of the first spin in four different cases. Panel (a) corresponds to relatively weak system-bath coupling $\lambda/\Omega=1/3$ and weak nearest-neighbor spin coupling; in panel (b), the spin-spin coupling is increased; in panel (c), the system-bath coupling is increased; and in panel (d), both couplings are strong. In all four cases, the CDH representation of $M=3$ converges more rapidly than the corresponding calculation of $M=3$ RC. Consistent with the single-spin results of Sec.~\ref{sec:TLS}, at weak system--bath coupling the CDH accurately reproduces both the phase and oscillation frequency of the dynamics, while the relaxation dynamics is well captured at both weak and strong coupling.

Comparing the four cases further shows that convergence of both the RC and CDH calculations is affected more strongly by increasing the system-bath coupling than by increasing the direct spin-spin interaction. This behavior can be understood from the structure of the transformed Hamiltonian: the $XX$ interaction is internal to the spin system and is unaffected by the RC mapping and the polaron transformation, whereas the $\hat{\sigma}^{z}_{\rm CDH}$ terms are dressed by the collective environmental mode. Increasing the system-bath coupling therefore enhances spin-RC hybridization and requires a larger RC subspace for convergence. A key observation is that accurate dynamics is already obtained with the CDH using only $M=3$ levels, resulting in only a modest increase of the Hilbert space of the system.

\begin{figure*}[ht]
    \centering
    \includegraphics[width=0.95\linewidth]{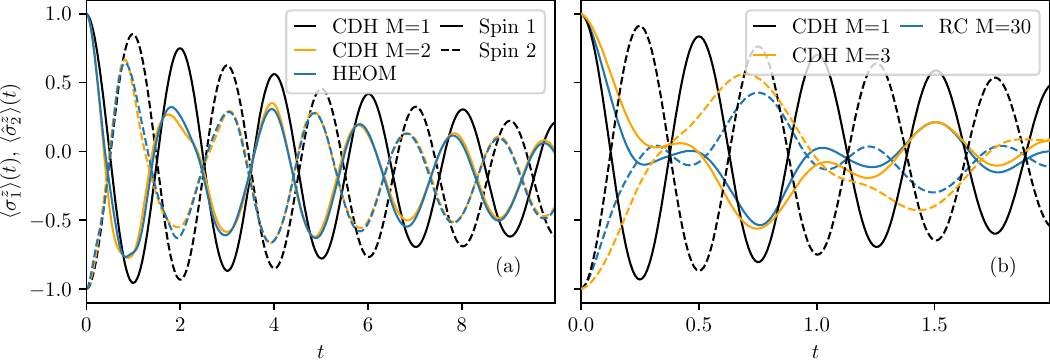}
\caption{Synchronization dynamics of two spins coupled through a common bosonic bath, initialized in the state $\ket{\uparrow\downarrow}$. The magnetizations $\braket{\hat{\sigma}^z_1}$ and $\braket{\hat{\sigma}^z_2}$ are shown by solid and dashed lines, respectively. In both panels, $\Delta_2=0.9\Delta_1$, $\gamma=0.05$, and $\beta=1/\Delta_1$. 
(a) $\Omega=8\Delta_1$ and $\lambda=2.5\Delta_1$. 
(b) $\Omega=8\Delta_1$ and $\lambda=5\Delta_1$. 
In panel (b), converged RC dynamics with $M=30$ are used as the reference since HEOM calculations did not converge even with a large number of auxiliary density operators.}
    \label{fig:spin_sync}
\end{figure*}

In Fig.~\ref{fig:spin_sync}, we examine the synchronization dynamics of two spins, initially prepared in the state $\ket{\uparrow\downarrow}$, at two different system-bath coupling strengths. In this case, the spins do not interact coherently-directly. We find that as in the single-spin case, at weak coupling the dynamics are captured accurately with a small CDH truncation (panel (a)): While the $M=1$ approach gets close to the synchronization frequency, it is evident by the later time in the simulation that both the frequency and the amplitude of the oscillations are overestimated. In contrast, already at $M=2$, the CDH method accurately reproduces both the amplitude and frequency of the oscillations, and thus the emergence of synchronized dynamics.

Panel (b) presents the dynamics at stronger system-bath coupling, where convergence requires a larger truncation. Here, simulations with $M=1$ fail to capture the dynamics \cite{brenesBathinduced2024}. In this regime, HEOM calculations (as implemented by QuTiP) suffer from numerical instability; therefore, the RC dynamics with $M=30$ was used as a numerical reference, after verifying convergence with respect to the RC truncation. CDH results systematically approach this reference as $M$ increases, demonstrating that the method can capture bath-induced synchronization dynamics in the strong-coupling regime.
Significantly, at this stronger coupling, the dynamics becomes more complex, with multiple frequencies contributing to the evolution of magnetization and synchronization consequently being partially lost. Nevertheless, even at low truncation with $M=3$, the CDH method captures the general envelope of the dynamics.

\subsubsection{Dicke Model}
\label{sec:collective}

As the above sections have shown, the CDH method can capture the key details of open system dynamics with a limited addition to the system size. This makes it convenient to use for larger systems, allowing for the study of collective behavior at strong coupling. 

A simple model which has inspired much theoretical interest for its collective effects is the Dicke model. The model, $N$ atoms interacting with a single electromagnetic mode (i.e. a cavity), was shown by Dicke in 1954 to exhibit a phenomenon called superradiance, where the atoms --if collectively excited---emit a single pulse of coherent light with intensity $I$ scaling like $I\propto N^2$ \cite{dickeCoherence1954}. 
The model was later expanded by Hepp and Lieb, who showed that in the thermodynamic limit ($N\rightarrow\infty$) the model exhibits a phase transition: At coupling below a critical coupling strength $\lambda_c$ the equilibrium mode population $\braket{\hat a^\dagger\hat a}$ is 0. For coupling greater than $\lambda_c$,
the mode population scales with $\lambda$ 
\cite{heppSuperradiant1973, kirtonIntroduction2019}. 
This result also holds in the case of the open Dicke model, where the mode suffers losses to an environment. Although the loss shifts the critical coupling, it does not change the overall phase transition \cite{kirtonIntroduction2019, brangeLeeYang2024}. 

We extend Eqs. \eqref{eq:2_spin_ham} and \eqref{eq:2_spins_trans} to $N$ spins, taking $\Delta_i$ to be identical for all spins. We also set the $XX$ coupling to 0, and rescale the system-bath coupling by $1/\sqrt{N}$. This results in the CDH Hamiltonian
\begin{equation}
    \hat H^S_\text{cdh} = \Delta\sum_{i=1}^N(\hat\sigma_i^z)_\text{cdh} + \Omega\hat a^\dagger \hat a - \frac{\lambda^2}{\Omega N}\sum_{i\neq j}\hat\sigma_i^x\hat\sigma_j^x.
\end{equation} 
Notice that the transformation introduces an all-to-all $XX$ coupling between spins, reflecting their effective interaction via the bath mode. 
The new system operator that couples to the bath is
\begin{equation}
    \hat S_\text{cdh} = \hat a +\hat a^\dagger -\frac{2\lambda}{\Omega\sqrt{N}} \sum_{i=1}^N\hat\sigma^x_i.
\end{equation}

\begin{figure*}[ht]
    \centering
\includegraphics[width=0.95\linewidth]{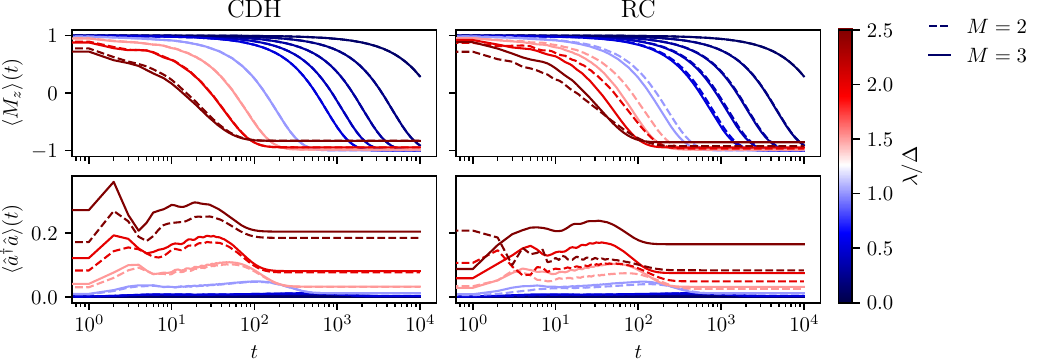}
    \caption{Average magnetization and cavity mode photon number for 5 spins coupled to a common bath using CDH and RC methods. All  spins have the same spin splitting $\Delta$ with initial state $\ket{\uparrow\uparrow\uparrow\uparrow\uparrow}$. Cavity and bath parameters are  $\Omega=8\Delta$, $\gamma=0.05$, $M=2$, 3. The bath and RC are initialized to a vacuum state, corresponding to $T=0\Delta$.}
    \label{fig:spin_chain}
\end{figure*}


In Fig.~\ref{fig:spin_chain}, we present the dynamics of the net magnetization,
$\hat{M}_z = \frac{1}{N}\sum_{i=1}^N \hat{\sigma}_i^z$ (top), and the cavity photon number, $\hat{a}^\dagger\hat{a}$ (bottom), for truncations $M=2$ and $M=3$. The spins are initialized in the fully excited state, the RC in its vacuum state, and
the bath at zero temperature. From a computational perspective, extending the
system by an RC truncated to $M=2$ or $M=3$ levels is equivalent, in terms of
Hilbert-space dimension, to adding just a single spin-$1/2$ or spin-1 degree of
freedom, respectively.

The CDH dynamics show relatively little change with increasing truncation from $M=2$ to $M=3$, indicating rapid convergence. In contrast, RC results exhibit a stronger dependence on $M$. As the truncation is increased, both the peaks in the cavity population and
the decay rate of the magnetization evolve toward the corresponding CDH
results.
Thus, for truncations of only $M=2$ and $M=3$ levels, the CDH provides a more accurate description of both spin and cavity dynamics than the conventional RC approach.




The dynamics of $\langle \hat{a}^\dagger \hat{a}\rangle$ in
Fig.~\ref{fig:spin_chain} display pronounced peaks associated with collective emission into the cavity mode. The population of this truncated mode provides a useful probe of the dynamics of the strongly coupled environmental degrees of freedom.
Furthermore, the long-time cavity population increases with increasing
$\lambda$, consistent with growing cavity occupation associated in the thermodynamic limit with the
transition toward the superradiant phase of the Dicke model.
However, the finite system considered here does not yet permit identification of a thermodynamic superradiant phase transition.

We emphasize that the calculations presented here for a spin ensemble focus on the nonresonant light-matter regime, with the spin splitting significantly smaller than the
cavity frequency, $\Delta \ll \Omega$. 
An important future direction is to investigate the near-resonant regime, where stronger hybridization between the spin and cavity degrees of freedom may lead to richer dynamics.
Furthermore, it is possible to use the CDH method and readily examine other cavity observables beyond the photon population, including the cavity displacement and quadrature
fluctuations, which can provide information about field displacement and
squeezing. 

\section{Summary}
\label{sec:summ}

We investigated the dissipative dynamics of open quantum systems using the compact dressed Hamiltonian framework. The method combines reaction-coordinate mapping with a polaron transformation. Truncating the transformed reaction-coordinate space yields a compact effective representation of the model Hamiltonian, which incorporates strong coupling into the parameters of the system Hamiltonian. At the same time, under the mapping, the coupling to the residual environment becomes sufficiently weak to be treated using a Markovian master equation of Redfield form.

We performed the CDH mapping and studied the dissipative dynamics of three models: the spin-boson model, two coupled spins, and spin ensembles. We found that the CDH method consistently provided a more accurate description than a direct truncation of the reaction-coordinate Hamiltonian at the same Hilbert-space dimension. Benchmarking against numerically exact HEOM simulations showed that the improvement is particularly pronounced at strong coupling, where the hybridized basis generated by the polaron transformation captured interactions  more efficiently. The CDH method accurately reproduced oscillation frequencies, relaxation behavior, and long-time populations using only a small number of reaction-coordinate levels. Although convergence becomes more demanding close to resonance and as dimensionless coupling $\lambda/\Omega$ increases, the CDH representation generally approaches the converged result more rapidly than the bare reaction-coordinate representation.

A central advantage of the CDH framework, when combined with a low-order quantum master equation such as the Redfield equation, is that it provides direct access to the spectrum and eigenmodes of the Liouvillian superoperator. The low-lying Liouvillian eigenvalues, which govern the slow relaxation dynamics and the approach to the steady state, are significantly more robust to Hilbert-space truncation than those obtained from the conventional direct RC representation. The CDH framework thus enables accurate extraction of relaxation timescales and identification of the dominant dissipative modes without requiring expensive simulations.
%
%
%
Once convergence with respect to the RC truncation is achieved in the CDH
representation, the remaining discrepancies with numerically exact simulations can be attributed primarily to the weak coupling and Markov approximations used to
describe the coupling to the residual bath, rather than to the CDH truncation itself. 

Future work could combine the CDH representation with non-Markovian or higher-order quantum master equations, extending the framework applicability to lower temperatures and to environments with more general spectral densities. 
A natural extension of the present framework is to consider structured environments with multiple characteristic frequencies, which could be accommodated by introducing several reaction-coordinate modes. Such a generalization would highlight a key advantage of the CDH approach over alternative quantum master-equation methods, such as polaron-frame QMEs, which are typically restricted to specific classes of bath spectral densities, such as of Ohmic form. The ability to construct compact dressed Hamiltonians for environments with general spectral densities would significantly broaden the applicability of the CDH formalism.

In sum, the CDH representation provides an accurate, computationally
efficient, and physically transparent framework for simulating strongly
coupled open quantum-system dynamics. Its potential to provide direct access to the relaxation timescales and modes is particularly appealing for chemical reaction dynamics and quantum information applications. 

\begin{acknowledgments}
D.S. acknowledges support from an NSERC Discovery Grant.
The authors thank M. Thoss for helpful discussions and S. Gatto for HEOM simulations used in panels (a3), (b3), and (c3) of Fig. \ref{fig:conv} and subsequent analysis.
The authors acknowledge J. Garwo{\l}a for the discussions that motivated and advanced  this work.
\end{acknowledgments}

The data presented in this article are available at the Github repository in Ref. \cite{gilesZgilesuoft2026}.

\begin{widetext}
   
\appendix

\section{Additional Simulations}\label{s:T_and_g}

\begin{figure}[htbp]
    \centering
\includegraphics[width=0.95\linewidth]{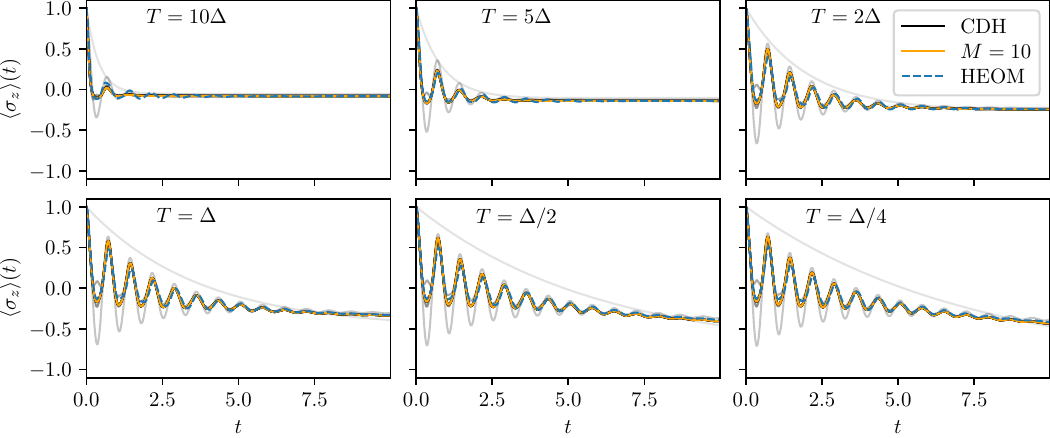}
    \caption{Spin-boson magnetization dynamics for $\Omega=2\Delta$, $\lambda=\Delta$, $E=0$, with decreasing temperature $T=10\Delta$, $5\Delta$, $2\Delta$, $\Delta$, $\Delta/2$ and $\Delta/4$.
    We present CDH results with increasing $M=1-9$ (grey), CDH at $M=10$ (yellow) and HEOM simulations (dashed blue).
    }
    \label{fig:var_T}
\end{figure}

\begin{figure}[htbp]
    \centering
\includegraphics[width=0.95\linewidth]{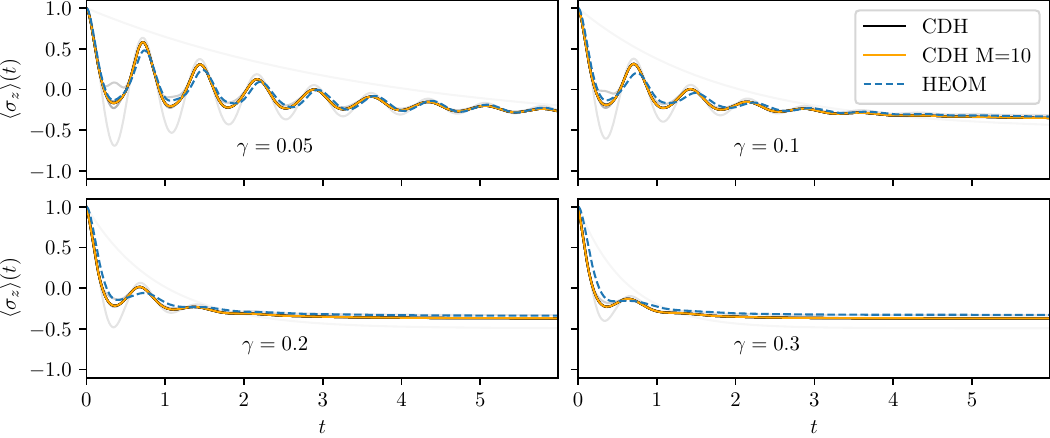}
    \caption{Spin-boson magnetization dynamics for $\Omega=9\Delta$, $\lambda=5\Delta$, $\beta=1/\Delta$, $E=0$ with increasing spectral width $\gamma=$ 0.05, 0.1, 0.2, and 0.3.
       We present CDH results with increasing $M=1-9$ (grey), CDH at $M=10$ (yellow) and HEOM simulations (dashed blue).
    }
    \label{fig:var_gam}
\end{figure}
In Figs.~\ref{fig:var_T} and \ref{fig:var_gam}, we compare the dynamics of CDH with HEOM while varying the temperature $T$ and the width parameter $\gamma$ in the spectral density function, corresponding to the parameter sweeps shown in Figs.~\ref{fig:meth_temp_gamma}(c) and \ref{fig:meth_temp_gamma}(d), respectively. 
The closest agreement between CDH and HEOM is observed at intermediate temperatures, $T\sim\Delta$, and for small values of $\gamma$. At higher temperatures, increased thermal occupation of the RC makes higher excited states relevant, and therefore a large CDH truncation is required. At low temperatures, on the other hand, memory effects become increasingly important, reducing the accuracy of the Markovian Redfield QME used to describe the residual bath. 

Increasing $\gamma$ also reduces the agreement between the CDH and HEOM dynamics. A broader spectral density modifies the coupling of the RC to the residual bath. It enhances both dissipative effects and the relevance of bath memory, thus increasing the error associated with the weak-coupling Markovian treatment of the residual environment. Consequently, as $\gamma$ increases, a more careful treatment of the residual bath is required, beyond the Markovian Redfield approximation.


\begin{figure}[htbp]
    \centering
\includegraphics[width=0.85\linewidth]{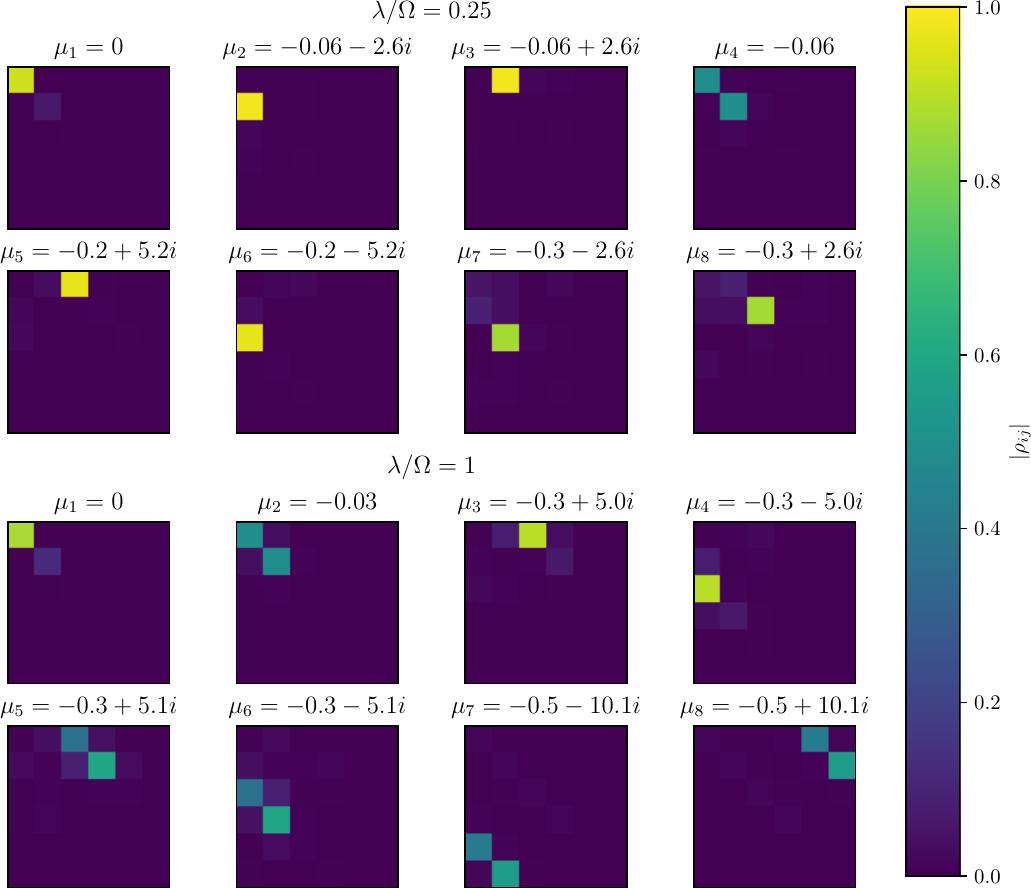}
    \caption{First eight eigenmodes and eigenvalues $\mu_i$ (ordered by the real part of their eigenvalues) of the Redfield superoperator in its energy eigenbasis for the spin-boson model, Eq \eqref{eq:spin_boson_cdh}. The parameters used are $M=3$ (for an overall matrix size of $6\times6$), $E=\Delta$, $\Omega=5\Delta$, $\beta=1/\Delta$, $\gamma=0.05$,  $\lambda = \Omega/4$ and $\lambda=\Omega$.}
    \label{fig:eig_vecs}
\end{figure}
\section{Eigenmatrices of the CDH Transformed Spin-Boson Model}\label{s:eigenmats}

The eigenmatrices of the Redfield tensor of the CDH transformed spin-boson model, in the energy eigenbasis of the Hamiltonian, with $M=3$ are plotted in Fig. \ref{fig:eig_vecs} at $\lambda/\Omega=0.25$ and 1, corresponding to the dashed black lines in Fig. \ref{fig:eig_v_m_bias}.

The map shows that the first eigenmatrices corresponding to populations are associated with $\mu_4$ for $\lambda/\Omega=0.25$ and $\mu_2$ for $\lambda/\Omega = 1$. When examined  with Figs. \ref{fig:eig_v_m_bias} and \ref{fig:mu_2}, we see that this corresponds to the extracted smallest non-zero real eigenvalues, confirming that Fig. \ref{fig:mu_2} does manifest the population relaxation rate, that is, the eigenvalue corresponding to the first non-steady state population eigenmode.


\section{Analysis of Liouvillian Eigenvalues of the Spin-Boson model at $M=1$}\label{s:eigenvalues}

\subsection{Unbiased model, $E=0$}
We study the dynamics of the CDH and RC Hamiltonians using the Redfield master equation, given by \cite{nitzanChemical2006}
%
\begin{multline}
    \odv{\rho_{mn}(t)}{t} = -i\omega_{mn}\rho_{mn}(t) - \sum_{ij}\left[R_{mi,ij}(\omega_{ji})\rho_{jn}(t)+R_{nj,ji}^*(\omega_{ij})\rho_{mi}(t) - \left(R_{jn,mi}(\omega_{im})+R^*_{im,nj}(\omega_{jn})\right)\rho_{ij}(t)\right].
\end{multline}
%
Here, $\omega_{mn}$ correponds to the Bohr frequency between levels $m$ and $n$.
The Redfield tensor is given by
\begin{equation}
    R_{ij,lm}(\omega) = S_{ij}S_{lm}\frac{\Gamma(\omega)}{2},
\end{equation}
where $\Gamma(\omega)$ is the real part of the correlation function $\braket{\hat B(t)\hat B(0)}$. Here, $\hat B(t)$ is the bath operator in the interaction picture. In terms of the spectral density, it can be shown that
\begin{equation}
    \Gamma(\omega) = \begin{cases}
        2J(|\omega|)n_B(|\omega|) & \omega < 0 \\
        2J(\omega)(1+n_B(\omega)) & \omega > 0 \\
        \lim_{\omega\rightarrow0}2J(\omega)n_B(\omega) & \omega = 0
    \end{cases}.
\end{equation}
For the CDH Hamiltonian $J(\omega) = \gamma\omega$ (taking $\Lambda\rightarrow\infty$ in Eq. \eqref{eq:ohmic}), and
\begin{equation}
    \Gamma(\omega) = \begin{cases}
        \gamma\omega(\coth(\frac{\beta\omega}{2})+1) & \omega \neq 0 \\
        \frac{2\gamma}{\beta} & \omega = 0
    \end{cases}.
\end{equation}
The CDH Hamiltonian of Eq. \eqref{eq:spin_boson_cdh} with $E=0$,  truncated to $M=1$, has the form
\begin{equation}
    H_\text{eff} = \tilde\Delta\hat\sigma_z - 2\epsilon\tilde{\sum_k} f_k\hat\sigma_x(\hat b_k^\dagger + \hat b_k) + \tilde{\sum_k}  \omega_k\hat b_k^\dagger\hat b_k
\end{equation}
where $\tilde\Delta = \Delta e^{-2\epsilon^2}$. Let index 1 denotes the ground state of the system, and index 2 denotes the excited state of the system. Then, transition energies are given by
\begin{equation}
    \omega_{11}=\omega_{22} = 0,\,\omega_{21}=-\omega_{12} = 2\tilde\Delta,
\end{equation}
Since $\hat S=\hat\sigma_x$, $S_{11}=S_{22}=0$ and  $R_{ij,lm} = 0$ if $i=j$ or $l=m$. Altogether, we obtain the equations of motion for the density matrix from the Redfield QME,
\begin{align}
    \dot\rho_{11}(t) &= 4\epsilon^2\left[\Gamma(2\tilde\Delta)\rho_{22}(t)-\Gamma(-2\tilde\Delta)\rho_{11}(t)\right] \\
    \dot\rho_{22}(t) &= 4\epsilon^2\left[\Gamma(-2\tilde\Delta)\rho_{11}(t)-\Gamma(2\tilde\Delta)\rho_{22}(t)\right] \\
    \dot\rho_{12}(t) &= 2\left[i\tilde\Delta-\epsilon^2[\Gamma(2\tilde\Delta)+\Gamma(-2\tilde\Delta)]\right]\rho_{12}(t)+2\epsilon^2[\Gamma(2\tilde\Delta)+\Gamma(-2\tilde\Delta)]\rho_{21}(t) \\
    \dot\rho_{21}(t) &= -2\left[i\tilde\Delta+\epsilon^2[\Gamma(2\tilde\Delta)+\Gamma(-2\tilde\Delta)]\right]\rho_{21}(t) +2\epsilon^2[\Gamma(2\tilde\Delta)+\Gamma(-2\tilde\Delta)]\rho_{12}(t).
\end{align}
We now define
\begin{equation}
    \tilde\Gamma = \Gamma(2\tilde\Delta)+\Gamma(-2\tilde\Delta) = 4\gamma\tilde\Delta\coth(\beta\tilde\Delta).
\end{equation}
Rewriting the equations of motion as a matrix acting on a vectorized density matrix gives
\begin{equation}
    \begin{pmatrix}
        \dot\rho_{11} \\ \dot\rho_{22} \\ \dot\rho_{12} \\ \dot\rho_{12}
    \end{pmatrix} = \begin{pmatrix}
        -4\epsilon^2\Gamma(-2\tilde\Delta) & 4\epsilon^2\Gamma(2\tilde\Delta) & 0 & 0 \\
        4\epsilon^2\Gamma(-2\tilde\Delta) & -4\epsilon^2\Gamma(2\tilde\Delta) & 0 & 0 \\
        0 & 0 & 2i\tilde\Delta-2\epsilon^2\tilde\Gamma & 2\epsilon^2\tilde\Gamma \\
        0 & 0 & 2\epsilon^2\tilde\Gamma & -2i\tilde\Delta-2\epsilon^2\tilde\Gamma
    \end{pmatrix}\begin{pmatrix}
        \rho_{11} \\ \rho_{22} \\ \rho_{12} \\ \rho_{21}
    \end{pmatrix}.
\end{equation}
Notice thate that the populations and coherences are split into separate sectors. The eigenvalues of the population sectors are
\begin{equation}
    \mu_1 = 0,\,\mu_2=-4\epsilon^2\tilde\Gamma.
    \label{eq:Epop}
\end{equation}
The eigenvalues of the coherence sector are
\begin{equation}
    \mu_{3(4)} = -2\epsilon^2\tilde\Gamma\pm2\sqrt{\epsilon^4\tilde\Gamma^2-\tilde\Delta^2}
    \label{eq:Ecoh}
\end{equation}
Since 
$\tilde\Delta\rightarrow0$ when $\epsilon\gg1$, in this limit $\tilde\Gamma\rightarrow\frac{4\gamma}{\beta}$ and the eigenvalues are
\begin{equation}
    \mu_1=\mu_3=0,\,\,\,\mu_2=\mu_4=-\frac{16\gamma}{\beta}\epsilon^2.
\end{equation}
%
\begin{figure*}[ht]
    \centering
\includegraphics[width=0.95\linewidth]{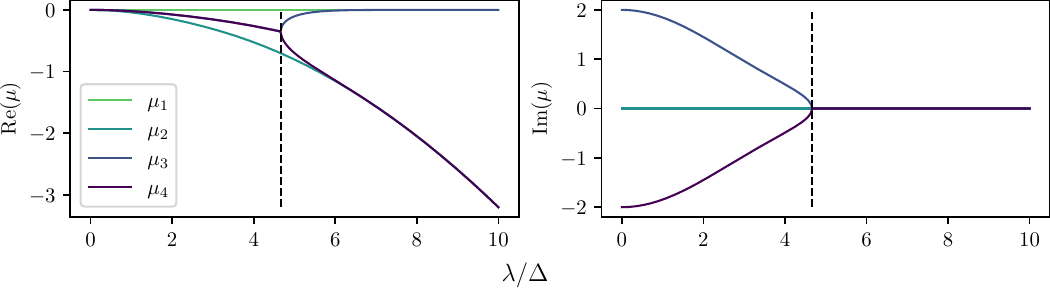}
    \caption{Eigenvalues of the Liouvillian of the CDH spin-boson model with $M=1$ based on Eqs. (\ref{eq:Epop})-(\ref{eq:Ecoh}). The dashed black line marks the value of the coupling energy $\lambda$ where the real parts of $\mu_3$ and $\mu_4$ bifurcate and their imaginary parts go to zero. Parameters used are $\Omega=5\Delta$, $\gamma=0.05$, $\beta=1/\Delta$.}
    \label{fig:ana_eig}
\end{figure*}

The real and imaginary parts of the eigenvalues are plotted in Fig. \ref{fig:ana_eig} as a function of $\lambda$, corresponding to the numerically obtained eigenvalues in the top left panel of Fig. \ref{fig:eig_v_m_bias} with $E=0$ and $M=1$. Note that the real parts of $\mu_3$ and $\mu_4$ split at $\lambda/\Delta \approx 4.66$ ($\lambda/\Omega\approx0.932$)---marked by the dashed black line---while the imaginary parts go to 0. This is the point in Fig. \ref{fig:mu_2} where the smallest magnitude, non-zero, real eigenvalue stops corresponding to the relaxation rate of the first population excitation.

\subsection{Biased model, $E\neq0$}

The coefficients for the Redfield tensor are obtained from SymPy. Letting $U = \sqrt{\tilde\Delta^2+E^2}$, and redefining
\begin{equation}
    \tilde\Gamma = \Gamma(2U)+\Gamma(-2U) = 4\gamma U\coth(\beta U),
\end{equation}
the equations of motion are
\begin{multline}
    \begin{pmatrix}
        \dot\rho_{11} \\ \dot\rho_{22} \\ \dot\rho_{12} \\ \dot\rho_{21}
    \end{pmatrix} \\= \begin{pmatrix}
        -\frac{4\tilde\Delta^2}{U^2}\epsilon^2\Gamma(-2U) & \frac{4\tilde\Delta^2}{U^2}\epsilon^2\Gamma(2U) & -\frac{8\gamma E\tilde\Delta}{\beta U^2}\epsilon^2 & -\frac{8\gamma E\tilde\Delta}{\beta U^2}\epsilon^2 \\
        \frac{4\tilde\Delta^2}{U^2}\epsilon^2\Gamma(-2U) & -\frac{4\tilde\Delta^2}{U^2}\epsilon^2\Gamma(2U) & \frac{8\gamma E\tilde\Delta}{\beta U^2}\epsilon^2 & \frac{8\gamma E\tilde\Delta}{\beta U^2}\epsilon^2 \\
        -\frac{4E\tilde\Delta}{U^2}\epsilon^2\Gamma(-2U) & \frac{4E\tilde\Delta}{U^2}\epsilon^2\Gamma(2U) & -\frac{2\tilde\Delta^2}{U^2}\epsilon^2\tilde\Gamma-\frac{16\gamma E^2}{\beta U^2}\epsilon^2+2iU & \frac{2\tilde\Delta^2}{U^2}\epsilon^2\tilde\Gamma \\
        -\frac{4E\tilde\Delta}{U^2}\epsilon^2\Gamma(-2U) & \frac{4E\tilde\Delta^2}{U^2}\epsilon^2\Gamma(2U) & \frac{2\tilde\Delta^2}{U^2}\epsilon^2\tilde\Gamma & -\frac{2\tilde\Delta^2}{U^2}\epsilon^2\tilde\Gamma-\frac{16\gamma E^2}{\beta U^2}\epsilon^2-2iU
    \end{pmatrix}\begin{pmatrix}
        \rho_{11} \\ \rho_{22} \\ \rho_{12} \\ \rho_{21}
    \end{pmatrix}.
\end{multline}
For small $\lambda$ and $E<\Delta$, as in the right hand side of Fig. \ref{fig:eig_v_m_bias}, the terms with factors  $\frac{E\tilde\Delta}{U^2}$ should be suppressed relative to the terms with factor $\frac{\tilde\Delta^2}{U^2}$. Therefore, the scaling of the population eigenvalue should be similar to the unbiased case---growing like $\lambda^2$---with the shifted energy scale $U$ playing the role of $\tilde\Delta$. In the large $\lambda$ limit $\tilde\Delta\rightarrow0$ while $U$ and consequently $\Gamma(\pm2U)$ and $\tilde\Gamma$ tend to constant values. Thus, the only terms that survive are the diagonal terms for $\rho_{12}$ and $\rho_{21}$, which go to $-\frac{16\epsilon^2\gamma}{\beta}\pm2iE$, respectively. As such, in this limit the population eigenvalues go to zero, while the real part of the coherences grows like $\lambda^2$, confirming the behavior seen in Figs. \ref{fig:eig_v_m_bias} and \ref{fig:mu_2}. In particular, this confirms the turnover in the population eigenvalue, which goes from scaling like $\lambda^2$ for a small lambda to approaching $0$ for a large $\lambda$. 

\end{widetext}

\bibliography{CDH_Dynamics_bibtex}
\end{document}